\documentclass[twoside,12pt]{article}
\usepackage{CJK}
\usepackage{indentfirst}
\usepackage{bm}
\usepackage{epsfig}
\usepackage{mathptmx}
\usepackage{amsmath}
\usepackage{float}
\usepackage{epstopdf}
\usepackage{url}
\usepackage{bm}
\usepackage{cite}
\usepackage{color}
\usepackage{soul}
\usepackage{color, xcolor}

\newcommand{\upcite}[1]{\textsuperscript{\textsuperscript{\cite{#1}}}}

\begin{document}


\thispagestyle{empty} \vspace*{0.8cm}\hbox
to\textwidth{\vbox{\hfill\huge\sf Commun. Theor. Phys.\hfill}}
\par\noindent\rule[3mm]{\textwidth}{0.2pt}\hspace*{-\textwidth}\noindent
\rule[2.5mm]{\textwidth}{0.2pt}


\begin{center}
\LARGE\bf Nonequilibrium kinetics effects in Richtmyer-Meshkov instability and reshock processes$^{*}$
\end{center}

\footnotetext{\hspace*{-.45cm}\footnotesize $^\dag$Corresponding author, E-mail: xu\_aiguo@iapcm.ac.cn }
\footnotetext{\hspace*{-.45cm}\footnotesize $^\ddag$Corresponding author, E-mail: wang\_lifeng@iapcm.ac.cn }

\begin{center}
\rm Yiming Shan$^{\rm a,b)}$, \ \ Aiguo Xu$^{\rm a,c,d)\dagger}$, \ \ Lifeng Wang$^{\rm a,c)\ddag}$,\ and \ Yudong Zhang$^{\rm e)}$
\end{center}

\begin{center}
\begin{footnotesize} \sl
\textit{}{National Key Laboroatory of Computational Physics, Institute of Applied Physics and Computational Mathematics, Beijing 100088, China}$^{\rm a)}$ \\
\textit{}{National Key Laboratory of Shock Wave and Detonation Physics, Institute of Fluid Physics, China Academy of Engineering Physics, Mianyang 621900, China.}$^{\rm b)}$ \\
\textit{Center for Applied Physics and Technology, MOE Key Center for High Energy Density Physics Simulations, College of Engineering, Peking University, Beijing 100871, China}$^{\rm c)}$ \\
\textit{State Key Laboratory of Explosion Science and Technology, Beijing Institute of Technology, Beijing 100081, China}$^{\rm d)}$ \\
\textit{School of Mechanics and Safety Engineering, Zhengzhou University, Zhengzhou 450001, China}$^{\rm e)}$ \\
\end{footnotesize}
\end{center}

\begin{center}
\footnotesize (Received XXXX; revised manuscript received XXXX)

\end{center}

\vspace*{2mm}

\begin{center}
\begin{minipage}{15.5cm}
\parindent 20pt\footnotesize
Kinetic effects in the inertial confinement fusion ignition process are far from clear.
 In this work, we study the Richtmyer-Meshkov instability (RMI) and reshock processes by using a two-fluid discrete Boltzmann method (DBM). The work begins from interpreting the experiment conducted by Collins and Jacobs [J. Fluid Mech. 464, 113-136 (2002)]. It shows that the  shock wave causes substances in close proximity to the substance interface to deviate more significantly from their thermodynamic equilibrium state.
The Thermodynamic Non-Equilibrium (TNE) quantities exhibit complex but inspiring kinetic effects in the shock process and behind the shock front.
 The kinetic effects are detected by two sets of TNE quantities. The first set includes
 $\left |\bm{\Delta} _{2}^{ {\rm{*}}}\right |$,
 $\left |\bm{\Delta} _{3,1}^{ {\rm{*}}}\right |$, $\left | \bm{\Delta} _{3}^{ {\rm{*}}}\right |$, and
  $\left |\bm{\Delta} _{4,2}^{ {\rm{*}}}\right |$, which correspond to the intensities of the Non-Organized Momentum Flux(NOMF), Non-Organized Energy Flux(NOEF), the flux of NOMF and the flux of NOEF.
 All four TNE measures abruptly increase in the shock process.
The second set of TNE quantities includes ${\dot{S} _{\mathrm{NOMF}} }$, ${\dot{S} _{\mathrm{NOEF}} }$ and ${\dot{S} _{\mathrm{sum}} }$, which denote the entropy production rates due to NOMF, NOEF and their summation, respectively.
The mixing zone is the primary contributor to ${\dot{S} _{\mathrm{NOEF}} }$, while the flow field region outside of the mixing zone is the primary contributor to ${\dot{S} _{\mathrm{NOMF}} }$.  Additionally, each substance exhibits different behaviors in terms of entropy production rate, and the lighter fluid has a higher entropy production rate than the heavier fluid.
\end{minipage}
\end{center}

\begin{center}
\begin{minipage}{15.5cm}
\begin{minipage}[t]{2.3cm}{\bf Keywords:}\end{minipage}
\begin{minipage}[t]{13.1cm}
Richtmyer-Meshkov instability; discrete Boltzmann method; nonequilibrium kinetics effects
\end{minipage}\par\vglue8pt

\end{minipage}
\end{center}

\section{Introduction}\label{sec:level1}
The Richtmyer-Meshkov instability\upcite{richtmyer1960taylor,meshkov1969instability} (RMI) is a fundamental hydrodynamic instability and is widely present in fields such as astrophysical applications\upcite{remington2006experimental} and inertial confinement fusion\upcite{nagel2022experiments,roycroft2022double,macphee2018hydrodynamic,sauppe2019using} (ICF). The interpretation and understanding of the physics of RMI are significant and necessary. In particular, RMI (and reshock) phenomena in ICF significantly affect the implosion performance, which has attracted much attention\upcite{li2021growth}. In addition, an increasing number of studies in recent years have shown that in ICF, due to the presence of high energy density particles and extreme and violent changes in physical quantities, there are strong nonequilibrium kinetic effects, which are nonnegligible and have an impact on the implosion process. In other fields, nonequilibrium behavioral characteristics induced by flow behaviors such as shock waves are also receiving increasing attention\upcite{xie2022chemical,li2021improved,liu2017molecular,qiu2020study,wu2022gaussian,yang2022spatio,jiang2019computation,CWF2016,li2015rarefied,LC2019,shan2006kinetic,meng2011accuracy,chen2022simulation,su2022temperature,guo2021progress}.


Rinderknecht $et$ $al.$\upcite{rinderknecht2018kinetic} noted that the systematic abnormalities in the National Ignition Facility (NIF) implosion dataset, including inferred missing energy in the hohlraum, drive asymmetry in near-vacuum hohlraums, low areal density and high burn-averaged ion temperatures compared with mainline models, and low ratios between the DD-neutron and DT-neutron yields and inferred Ti, suggest that kinetic physics may play a role. Shan $et$ $al$.\upcite{shan2018experimental} discussed that the kinetic shocks cause an anomalous large energy spread of the DD neutron signal and anomalous scaling of the neutron yield with the thickness of the CD layers, which are not explicable using hydrodynamic mechanics, and these findings were supported for the first time by experiments and simulations. Cai $et$ $al$.\upcite{cai2021hybrid} presented a new framework that combines kinetics and hydrodynamics to simulate shock waves and hydrodynamic instabilities in plasmas with high density. This hybrid approach preserves the kinetic effects of ions and their impacts on shock wave propagation, plasma interpenetration, and hydrodynamic instability. For two-fluid simulation, Zhang $et$ $al.$\upcite{zhang2020species} completed nonequilibrium molecular dynamics simulations of shock release in polystyrene (CH) under experimental conditions. Due to its lighter mass than carbon, hydrogen can stream out of the bulk of the CH foil, as observed in these simulations. This released hydrogen generates low-density plasmas in front of the in-flight shell, which is in agreement with the experimental findings. Zhang $et$ $al.$ noted that for ICF simulations, the kinetic effects of species separation are absent in single-fluid radiation-hydrodynamics codes. Although some studies have begun to pay attention to nonequilibrium kinetic effects, they are still very inadequate due to the lack of research methods, and the study of kinetic effects in ICF is still in its infancy.

The Richtmyer-Meshkov instability has been extensively studied\upcite{zhou2019turbulent,2017Rayleigh,zhou2017rayleigh,zou2020research} in theory\upcite{zhang1996analytical,valerio1999modeling,mikaelian1990rayleigh,zhang2018quantitative,liang2022phase}, numerical simulations\upcite{zhou2016asymptotic,lombardini2011atwood,tritschler2014evolution,ukai2011growth,olson2014comparison,hahn2011richtmyer,schilling2010high,yan2022effect,li2019role,thornber2012energy,li2018richtmyer,zou2019richtmyer,guan2017manipulation,sterbentz2022design}, and experiments\upcite{jones1997membraneless,balasubramanian2013experimental,jacobs2013experiments,zhai2016richtmyer,weber2012turbulent,vandenboomgaerde2014experimental,balasubramanian2012experimental,liu2018elaborate,luo2019nonlinear,collins2002plif,zou2017richtmyer,li2022instability,ding2017shock,lei2017experimental,luo2018long,liang2019richtmyer} for general conditions. The physical comprehension of RMI is becoming clearer. However, for general fluid cases, there are also significant nonequilibrium kinetic effects, and this part is missing for the majority of existing studies. For the RMI and reshock processes, in addition to the nonequilibrium effect due to material flow and mixing, shock wave effects are also present, which generate extremely strong nonequilibrium kinetic effects, and the two kinds of nonequilibrium effects interact with each other. Unfortunately, such effects cannot be or are not easily observed and described by traditional macroscopic fluid modeling methods, so studies of nonequilibrium effects present in RMI are scarce. Therefore, it is important to develop corresponding kinetic description methods and to detect and describe the kinetic effects therein.

The numerical exprimental study of complex flow includes three main steps: (i) physical modeling, (ii) discrete format selection/design, and (iii) numerical experiments and complex field analysis.
Discrete Boltzmann method (DBM) \upcite{xu2012lattice,Xu-Zhang-book-2022,gan2022discrete,zhang2022discrete,zhang2022non} is a modeling and analysis method based on the discrete Boltzmann equation and developed from the lattice Boltzmann method\upcite{xu2012lattice,succi2001lattice}. It focuses on the steps (i) and (iii), and leave the
 best discrete format design in
 step (ii) to computational mathematician.
As a modeling method, with increasing the degree of non-equilibrium/non-continuity, the DBM uses more kinetic moments of the distribution function $f$ to capture the main feature of the system.
As an analysis method, it provides various measures based on the non-conserved kinetic moments of $(f-f^{\mathrm{eq}})$ to detect, describe and exhibit complex the Thermodynamic Nonequilibrium (TNE) behavior, where $f^{\mathrm{eq}}$ is the corresponding equilibrium distribution function.
The DBM is a further development of the statistical physical phase space description method.
With the introduction of phase space based on the non-conserved kinetic moments of $(f- f^{\mathrm{eq}})$, we can further introduce the concept of behavior feature vectors by taking a group of behavior feature quantities as elements. For example, since any definition of the nonequilibrium strength depends on the research perspective, the description of the nonequilibrium behavior of the system obtained by using the nonequilibrium strength from a single perspective is incomplete, often losing much information. Thus, it is helpful to introduce the nonequilibrium strength vector, each element of which describes the degree/extent/intensity of system deviation from equilibrium from its own perspective, to describe the nonequilibrium state of the system from multiple perspectives\upcite{Xu-Zhang-book-2022}.

Since there are infinite nonconservative moments of the distribution functions, each independent component of any nonconservative moment can provide a corresponding perspective of the nonequilibrium strength. In addition, the time or space change rate of any physical quantity, such as the density, temperature, flow rate, and pressure, can provide a corresponding perspective of the nonequilibrium strength. Other parameters, such as the Knudsen numbers corresponding to different interfaces and the relaxation time of any nonconservative moment, can provide the nonequilibrium strength from the corresponding perspective. Therefore, the nonequilibrium strength vector constructed based on the nonequilibrium strengths from several different perspectives is still only one perspective for describing the nonequilibrium behavior of the system. However,
compared with the description based on the nonequilibrium strength from a single perspective, the nonequilibrium strength vector provides a much more accurate description.
It should also be mentioned that what DBM includes are a series of physical constraints on the discrete format, instead of the specific discrete format itself. The best discretization scheme is an open research subject of computational mathematics.

The physical function of DBM is roughly equivalent to a set of Extended Hydrodynamic Equations (EHE), where ``extended" means that, besides the evolution equations of conserved kinetic moments of the distribution function $f$, evolution equations of some most relevant nonconserved kinetic moments are included. The neccessity of the extended part of the EHE increases with increasing the Knudsen(Kn) number.
Therefore, being able to recover the hydrodynamic equations in corresponding level is only one part of the physical function of DBM. To which order of TNE the DBM should include depends on the specific situation. The simplest one that meets the needs is naturally the first choice.

DBM has brought about a series of new insights in several fields, such as for multiphase flow\upcite{zhang2020kinetic,sun2022,gan2022discrete}, rarefied gas flow\upcite{zhang2022non,zhang2018discrete}, combustion and detonation\upcite{Lin2016CNF,Lin2018CNF,shan2022discrete}, and hydrodynamic instabilities\upcite{Lai2016PRE,Gan2019PRE,chen2022}. In contrast to conventional CFD methods, the DBM is not restricted by continuous and near-equilibrium assumptions and can describe cases of high Kn numbers with strong TNE and noncontinuity effects.
When designing numerical examples in the literature\upcite{zhang2022discrete}, the description method of nonequilibrium strength vectors from different perspectives was adopted.

As mentioned above, kinetic effects and two-fluid numerical simulations are essential and realistic but lacking in fields such as ICF. In the present paper, we report on single-mode RMI and reshock processes numerically simulated with a two-fluid DBM\upcite{Lin2016CNF}. In the two-fluid DBM, each fluid substance has its own independent distribution function and control equation, and the model can provide nonequilibrium kinetics information for each substance. The numerical simulation conditions are the same as in Ref.\cite{latini2007high}, and the DBM results are in good agreement with the experimental results. Then, we investigate the nonequilibrium kinetic effects, molecular mixing fraction and entropy production rate in RMI and reshock and obtain some new insights. Investigating various nonequilibrium behaviors and kinetic effects in single-mode RMI and reshock problems by using the two-fluid DBM is expected to help improve the physical cognition of hydrodynamic instabilities. The remainder of the article is organized as follows. The two-fluid DBM is presented in Sec.2.  Section 3 presents the numerical simulation results obtained by the DBM and the discussion. The conclusions are presented in Sec.\ref{sec:conclusions}.

\section{The two-Fluid Discrete Boltzmann Method}\label{sec:method}
The two-fluid discrete Boltzmann-BGK equation can be written as\upcite{Lin2016CNF}:
\begin{equation}
\frac{{\partial f_i^\sigma }}{{\partial t}} + {v_{i\alpha }} \cdot \frac{{\partial f_i^\sigma }}{{\partial {r_{\alpha }}}} =  - \frac{1}{{{\tau ^\sigma }}}(f_i^\sigma  - f_i^{\sigma ,\mathrm{eq}}),
\end{equation}
where $\sigma $ is the index of the substance, with $\sigma {\rm{ = }}$ A or B in the two-fluid case. $i$ is the index of the discrete velocity, and $\alpha $ is the direction of the spatial coordinate system. $f_i^\sigma $ and $f_i^{\sigma ,\mathrm{eq}}$ are the discrete distribution function and the discrete equilibrium distribution function of substance A or B, respectively. ${v_{i\alpha }}$ is the discrete molecular velocity, ${r_{\alpha }}$ is the spatial coordinate, and $t$ is time. ${\tau ^\sigma }$ is the relaxation time. The number density, mass density and velocity of substance $\sigma $ are:
\begin{equation}
{n^\sigma } = \sum\limits_i {f_i^\sigma },
\end{equation}
\begin{equation}
{\rho ^\sigma }{\rm{ = }}{n^\sigma }{m^\sigma },
\end{equation}
\begin{equation}
{{\bm{u}}^\sigma } = \frac{{\sum\limits_i {f_i^\sigma {{\bm{v}}_i}} }}{{{n^\sigma }}},
\end{equation}
and the fluid exhibits overall number density $n$, mass density $\rho $ and velocity ${\bm{u}}$ values are:
\begin{equation}
n = \sum\limits_\sigma  {{n^\sigma }},
\end{equation}
\begin{equation}
\rho {\rm{ = }}\sum\limits_\sigma  {{\rho ^\sigma }},
\end{equation}
\begin{equation}
{\bm{u}} = \frac{{\sum\limits_\sigma  {{\rho ^\sigma }{{\bm{u}}^\sigma }} }}{\rho }.
\end{equation}

The substance $\sigma $ temperature ${T^\sigma }$ and the mixture temperature $T$ are:
\begin{equation}
{T^{\sigma }} = \frac{{2E_I^{\sigma {\rm{*}}}}}{{{n^\sigma }(D + {I^\sigma })}},
\end{equation}
\begin{equation}
T = \frac{{2E_I^*}}{{\sum\limits_\sigma  {{n^\sigma }(D + {I^\sigma })} }},
\end{equation}
where $E_I^{\sigma *}$ is the internal energy of substance $\sigma $.  $E_{I}^{\sigma *}=\frac{1}{2}m^{\sigma }\sum_{i}f_{i}^{\sigma }((\bm{v} _{i}-\bm{u}  )^{2}+\eta ^{\sigma 2}   )$ and $E_I^* = \sum\limits_\sigma  {E_I^{\sigma *}} $, ${\eta ^{\sigma 2} }$ is the energy of the extra degree of freedom. Because the definition of internal energy $E_I^{\sigma *}$
depends on the flow velocity chosen as a reference, the symbol ``$*$'' in the superscript of $E_I^{\sigma *}$ and $E_I^*$ means the macroscopic fluid velocity $\bm{u}$ is chosen to calculate the internal energy.  ${I^\sigma }$ is the extra degree of freedom, and $D$ is the spatial dimension.

Here, we use the one-step (relaxation collision) model in this paper to simulate the flow field evolution and the interactions of different substance particles, and
\begin{equation}
{f^{\sigma ,\mathrm{eq}}} = {\rho ^\sigma }{\left( {\frac{1}{{2\pi RT}}} \right)^{(D + {I^\sigma })/2}}\exp \left( { - \frac{{{c^2} + {\eta ^{\sigma 2}}}}{{2RT}}} \right).
\end{equation}
${c_\alpha }{\rm{ = }}{v_\alpha } - {u_\alpha }$ is the fluctuation of the molecular velocity in the $\alpha $ direction, and ${c^2} = {c_\alpha }{c_\alpha }$. $R$ is the perfect gas constant. In the one-step relaxation collision model of the two-fluid DBM, the interactions between the two substances particles are expressed in the ``equilibrium'' distribution functions of the each substance $f^{\sigma ,\mathrm{eq}}$: the density, velocity and temperature used to calculate the equilibrium distribution function of each substance are related to both substances. In detail, the density used to calculate the equilibrium distribution function of each substance is the substance A or B particle density ${\rho ^\sigma }$ , and the velocity and temperature are the mixture average velocity ${\bm{u}}$ and mixture average temperature $T$. In this way, the interactions between two substance particles are naturally contained in the two-fluid DBM.

It should be noted that the principle of DBM velocity space discretization is that the values of moments used to describe the system must remain unchanged after discretization.
The detection and description of nonequilibrium behavior are the core features of the DBM. The nonconservative moments of $f_i^\sigma  - f_i^{\sigma ,\mathrm{eq}}$ are used to characterize the degree of nonequilibrium in the flow field. We define:
\begin{equation}
\bm{\Delta}_n^{\sigma {\rm{*}}}  = {{\bm{M}}_n^{\rm{*}}}(f_i^\sigma - f_i^{\sigma ,\mathrm{eq}}),
\end{equation}
where ${{\bm{M}}_n^{\rm{*}}}(f_i^\sigma )$ and ${{\bm{M}}_n^{\rm{*}}}(f_i^{\sigma ,\mathrm{eq}})$ are the $n$-th order kinetic central moments of $f_i^\sigma $ and $f_i^{\sigma ,\mathrm{eq}}$. ${\bm{\Delta} _n^{\rm{*}}}$ is the TNE characteristic quantity. In general, for the DBM containning only the first order TNE, ${\bm{\Delta} _2^{\rm{*}}}$, ${\bm{\Delta} _{3,1}^{\rm{*}}}$, ${\bm{\Delta} _3^{\rm{*}}}$ and ${\bm{\Delta} _{4,2}^{\rm{*}}}$ are used. As for the subscripts, ``2" means the 2nd order tensor, ``3,1" means the first-order tensor contracted from the third-order tensor, and the rest are similar. Different from $E_I^{\sigma *}$ and $E_I^*$, the symbol ``$*$" in the superscript of ${\bm{\Delta} _n^{\rm{*}}}$ means the ${\bm{\Delta} _n^{\rm{*}}}$ is calculated by the kinetic central moments rather than the kinetic non-central moments of $f_i^\sigma $ and $f_i^{\sigma ,\mathrm{eq}}$. Through a comparison to the macroscopic fluid equations, we find that ${\bm{\Delta} _2^{\rm{*}}}$ (Non-Organized Momentum Flux, NOMF) represents the viscous stress and ${\bm{\Delta} _{3,1}^{\rm{*}}}$ (Non-Organized Energy Flux, NOEF) represents the heat flux. ${\bm{\Delta} _3^{\rm{*}}}$ and ${\bm{\Delta} _{4,2}^{\rm{*}}}$ are the non-organized  fluxes of ${\bm{\Delta} _2^{\rm{*}}}$ and ${\bm{\Delta} _{3,1}^{\rm{*}}}$, respectively. To provide a more convenient way to represent the degree of fluid deviation from equilibrium, we further define:
\begin{equation}\label{Eq:NEQ-1}
\left |\bm{\Delta} _2^{\sigma {\rm{*}}}\right |{\rm{ = }}\sqrt {\Delta _{2,xx}^{\sigma {\rm{*}} 2} + 2\Delta _{2,xy}^{\sigma {\rm{*}} 2} + \Delta _{2,yy}^{\sigma {\rm{*}} 2}} ,
\end{equation}
\begin{equation}\label{Eq:NEQ-2}
\left |\bm{\Delta} _{3,1}^{\sigma {\rm{*}}}\right |{\rm{ = }}\sqrt {\Delta _{3,1,x}^{\sigma {\rm{*}} 2} + \Delta _{3,1,y}^{\sigma {\rm{*}} 2}} ,
\end{equation}
\begin{equation}\label{Eq:NEQ-3}
\left |\bm{\Delta} _3^{\sigma {\rm{*}}}\right |{\rm{ = }}\sqrt {\Delta _{3,xxx}^{\sigma {\rm{*}} 2} + 3\Delta _{3,xxy}^{\sigma {\rm{*}} 2} + 3\Delta _{3,xyy}^{\sigma {\rm{*}} 2} + \Delta _{3,yyy}^{\sigma {\rm{*}} 2}} ,
\end{equation}
\begin{equation}\label{Eq:NEQ-4}
\left |\bm{\Delta} _{4,2}^{\sigma {\rm{*}}}\right |{\rm{ = }}\sqrt {\Delta _{4,2,xx}^{\sigma {\rm{*}} 2} + 2\Delta _{4,2,xy}^{\sigma {\rm{*}} 2} + \Delta _{4,2,yy}^{\sigma {\rm{*}} 2}} .
\end{equation}
Equations \eqref{Eq:NEQ-1}-\eqref{Eq:NEQ-4} give four perspectives to
characterize the nonequilibrium intensity/degree/extent of a state in terms of the distance of this state from the origin (equilibrium state) in the phase space based on the corresponding nonconservative kinetic moments ${{\bm{M}}_n^{\rm{*}}}(f_i^\sigma - f_i^{\sigma ,\mathrm{eq}})$.
As one research perspective, the total nonequilibrium intensities at arbitrary positions in the flow field are defined as ${\left |\bm{\Delta} _n^{\rm{*}}\right |}{\rm{ = }}\sum_{\sigma } {\left |\bm{\Delta} _n^{\sigma{\rm{*}}}\right |}$.
In addition, we can also define a total nonequilibrium degree in a certain flow field region,

\begin{equation}\label{Eq:NEQ-4b}
D^{\rm{*T}}_{n}
=
\sum_{i_x,i_y} \left | \bm{\Delta} _{n}^{*}   \right | ,
\end{equation}
where the summation is for the grid nodes.

Furthermore, a kind of non-dimensionalized and normalized nonequilibrium intensity quantities can be defined as below:
\begin{equation}\label{Eq:NEQ-5}
D^{\rm{*T}}_{2,\rm{nor}}
 = \frac{{D^{\rm{*T}}_{2} }}{{{D_{\mathrm{sum}}}}} ,
\end{equation}
\begin{equation}\label{Eq:NEQ-6}
D^{\rm{*T}}_{(3,1),\rm{nor}}
 = \frac{{D^{\rm{*T}}_{3,1} }}{{{D_{\mathrm{sum}}}}} ,
\end{equation}
\begin{equation}\label{Eq:NEQ-7}
D^{\rm{*T}}_{3,\rm{nor}}
= \frac{{D^{\rm{*T}}_{3} }}{{{D_{\mathrm{sum}}}}} ,
\end{equation}
\begin{equation}\label{Eq:NEQ-8}
D^{\rm{*T}}_{(4,2),\rm{nor}}
= \frac{{D^{\rm{*T}}_{4,2} }}{{{D_{\mathrm{sum}}}}} ,
\end{equation}
where
\begin{equation}\label{Eq:NEQ-8b}
{D_{\mathrm{sum}}} = \sqrt {
(D^{\rm{*T}}_{2})^2 + (D^{\rm{*T}}_{3,1})^2 + (D^{\rm{*T}}_{3})^2 + (D^{\rm{*T}}_{4,2})^2},
\end{equation}
corresponds to a total nonequilibrium intensity,
 and
\begin{equation}\label{Eq:NEQ-8c}
(D^{\rm{*T}}_{2,\mathrm{nor}})^2 + (D^{\rm{*T}}_{(3,1),\mathrm{nor}})^2 + (D^{\rm{*T}}_{3,\mathrm{nor}})^2 + (D^{\rm{*T}}_{(4,2),\mathrm{nor}})^2
=1.
\end{equation}
These new defined non-dimensionalized and normalized nonequilibrium quantities represent their respective proportions in the total nonequilibrium intensity ${D_{\mathrm{sum}}}$. This new approach to define nonequilibrium quantities can also be easily generalized to other nonequilibrium quantities. The two quantities, $D^{\rm{*T}}_{n}$ and the normalized nonequilibrium quantities $D^{\rm{*T}}_{n,\rm{nor}} $, investigate how the system deviates from equilibrium in the different perspectives.

All these perspectives are reasonable and have complementary effects.
For convenience of description, we can construct a nonequilibrium degree vector, each of its components represents the nonequilibrium degree from its own perspective. Two examples of nonequilibrium degree vector are as below:
\begin{equation}\label{Eq:NEQ-8d}
\mathbf{D}^{\rm{neq}} =
 \{
 D^{\rm{*T}}_{2}, D^{\rm{*T}}_{3,1}, D^{\rm{*T}}_{3}, D^{\rm{*T}}_{4,2},
 \}
\end{equation}
\begin{equation}\label{Eq:NEQ-8d}
\mathbf{D}^{\rm{neq}}_{\mathrm{nor}} =
  \{
 D^{\rm{*T}}_{2,\mathrm{nor}}, D^{\rm{*T}}_{(3,1),\mathrm{nor}}, D^{\rm{*T}}_{3,\mathrm{nor}}, D^{\rm{*T}}_{(4,2),\mathrm{nor}}
 \}.
\end{equation}

The entropy production rate is contributed by two parts: viscous stress ${\dot{S} _{\mathrm{NOMF}}^{\sigma} }$ and heat flux ${\dot{S} _{\mathrm{NOEF}}^{\sigma} }$:
\begin{equation}
{\dot{S} _{\mathrm{NOMF}}^{\sigma}} = \int { - \frac{1}{T}{\bm{\Delta} _2^{\sigma{\rm{*}}}}} :\nabla {\bm{u}}d{\bm{r}},
\end{equation}
\begin{equation}
{\dot{S} _{\mathrm{NOEF}}^{\sigma}} = \int {{\bm{\Delta} _{3,1}^{\sigma{\rm{*}}}} \cdot \nabla \frac{1}{T}d{\bm{r}}}.
\end{equation}
The total entropy production rate ${\dot{S} _{\mathrm{sum}}^{\sigma} }$ is:
\begin{equation}
{\dot{S} _{\mathrm{sum}}^{\sigma}} = {\dot{S} _{\mathrm{NOMF}}^{\sigma}} + {\dot{S} _{\mathrm{NOEF}}^{\sigma}}.
\end{equation}

\section{\protect\bigskip Results and Discussion}\label{sec:results}
\subsection{Nonequilibrium kinetic analysis of RMI and reshock}
In this paper, we use the DBM to reproduce the experimental results with the same simulation parameters as those mentioned in the literature \upcite{latini2007high} and analyze the nonequilibrium kinetic effects for this example.

\begin{figure*}[ht]
\centering
\includegraphics[height=7cm]{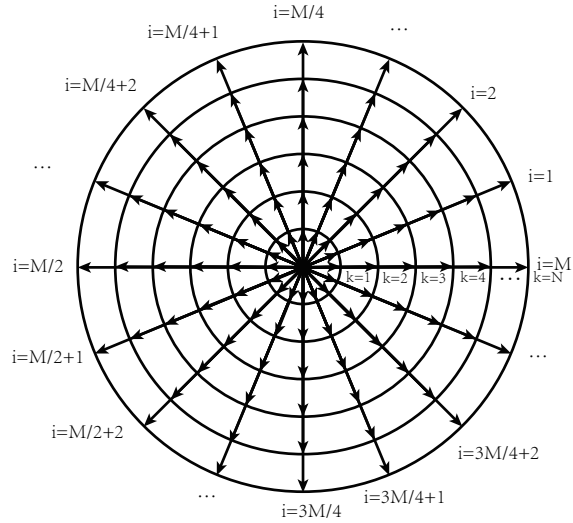}
\caption{Discrete velocity model.}
\label{Fig:DVM1}
\end{figure*}

\begin{figure*}[ht]
\centering
\includegraphics[height=7cm]{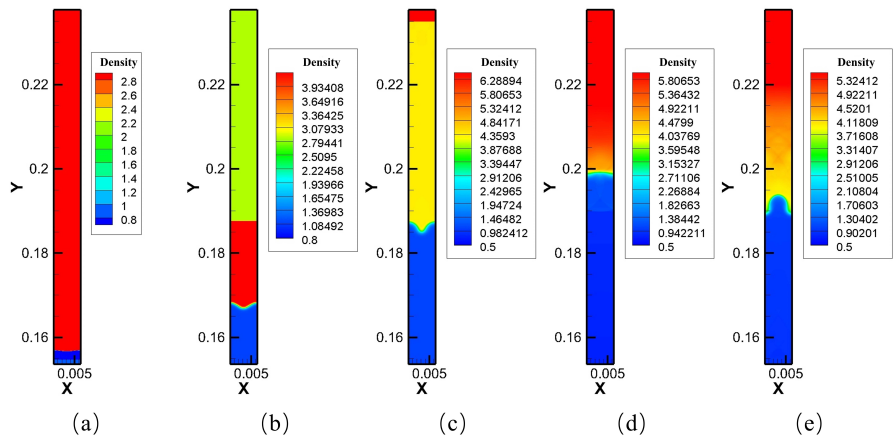}
\caption{Flow field evolution of single-mode RMI and reshock: (a) $t=0$, (b) $t=0.06$, (c) $t=0.162$, (d) $t=0.264$, and (e) $t=0.3$.}
\label{Fig:RMI1}
\end{figure*}

\emph{As a model construction and complex field analysis method, what the DBM presents are the basic constraints on the discrete formats.} The DBM itself does not give specific discrete formats. The specific discrete formats should be chosen according to the specific problem to be simulated.
The discrete velocity model used here is as follows:
\begin{equation}\label{Eq:DBM-6}
{{\bm{v}}_i} = {v_k}\left[ {\cos \left( {\frac{{i - 1}}{M} \cdot 2\pi } \right),\sin \left( {\frac{{i - 1}}{M} \cdot 2\pi } \right)} \right],
\end{equation}
where $k = 0,1, \cdots ,N $ and $i = 1,2, \cdots ,M $, with $N=4$ and $M=12$. $v_k$ is the value of the discrete velocity. The discrete velocity model diagram is shown in Fig. \ref{Fig:DVM1}.
The third-order Runge-Kutta scheme is used for temporal integration. The fifth-order weighted essentially non-oscillatory (WENO) scheme is used for calculating the spatial derivatives.
The flow field parameters are set as follows: space steps $dx = dy = 5\times 10^{-5} $, time step $dt = 3\times 10^{-6} $, and grid number $Nx \times Ny = 128 \times 4755$. The viscosity of the flow field is found to be small by evaluating the experimental and numerical simulation conditions mentioned in the Refs.\cite{collins2002plif,latini2007high}. In DBM, viscosity coefficient $\mu {\rm{ = }}\tau P$, where $P$ is the pressure. In order to match the experimental conditions, $\tau $ is chosen as small as possible, provided that numerical stability is satisfied. In this paper, $\tau {\rm{ = }}2\times 10^{-6} $.         Here, we define the substance below the interface as A and the substance above the interface as B, and $\rho _{A}<\rho _{B}$. Atwood number $At=(\rho_{B}-\rho_{A} )/(\rho_{B}+\rho_{A} )=0.6$. The specific heat ratio of substances A and B is $\gamma {\rm{ = }}1.25 $. The shock wave Mach number $Ma = 1.21$ and the shock conditions satisfy the Rankine-Hugoniot relations\upcite{zel2002physics}. The interface initial perturbation is $y = {A_0}\cos (kx)$, where perturbation wavenumber $k=2\pi /\lambda=981.75$, in which $\lambda=0.0064$ is the perturbation wavelength and amplitude $A_0=0.0002$. The shock is initiated at $y=0.00105$ in the lighter fluid and propagates in the positive direction of the $y$-axis. The light-heavy fluid initial interface is located at $y=0.0032$. Periodic boundary conditions are used for the left and right boundaries. The upper boundary has a reflection boundary condition, and the lower boundary has an inflow boundary condition. The parameters are selected considering grid independence, and Fig. \ref{Fig:RMI1} displays the results of the two-dimensional flow field simulation.

\begin{figure*}[ht]
\centering
\includegraphics[height=7cm]{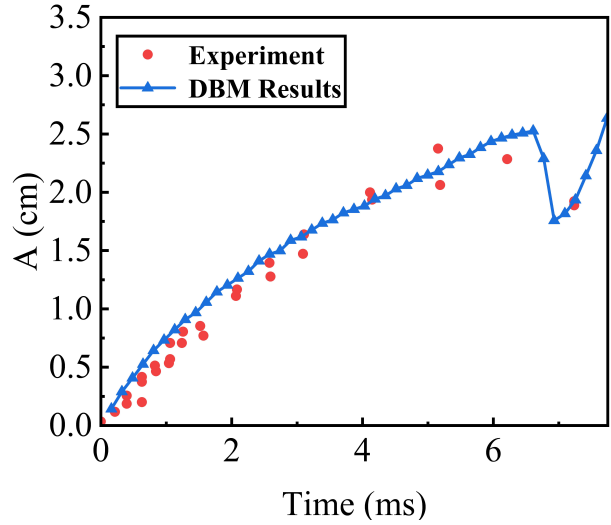}
\caption{Comparison of amplitude $A$ with time between DBM results and experimental data.}
\label{Fig:Z1}
\end{figure*}

Figure \ref{Fig:RMI1}(a) shows the initial flow field diagram, and the shock wave propagates along the positive $y$-axis. After passing through the material interface, RMI develops, and the shock front continues to propagate in the $y$-axis positive direction, which corresponds to Fig. \ref{Fig:RMI1} (b). When the shock front reaches the upper boundary, a reflected shock wave is produced and propagates in the negative $y$-axis direction, as shown in Fig. \ref{Fig:RMI1}(c). The reflected shock then reshocks the interface, and since this reflected shock wave is incident from the heavier fluid to the lighter fluid, the RMI interface is reversed (corresponding to Fig. \ref{Fig:RMI1} (d)). As a result of the reshock, a transmitted shock propagates in the negative $y$-axis direction, while a reflected shock starts to propagate upward, and the reversed RMI interface continues to evolve, as shown in Fig. \ref{Fig:RMI1} (e). Since the DBM results are all dimensionless, we recover the magnitude of the amplitude $A$ of RMI and time $t$ for comparison with the results in Refs.\cite{collins2002plif,latini2007high}, and the details on recovering the magnitude in the DBM can be found in Ref.\cite{zhang2019entropy}. Figure \ref{Fig:Z1} shows the comparison between the experimental results mentioned in Refs.\cite{collins2002plif,latini2007high} with the DBM simulation results after recovering the magnitude. The DBM results are in good agreement with the experimental results, which indicates that the DBM including only the first order TNE, the time step, space step, discrete velocity model and other parameter settings in the DBM program meet the needs of the current study.

\begin{figure*}[ht]
\centering
\includegraphics[height=6.5cm]{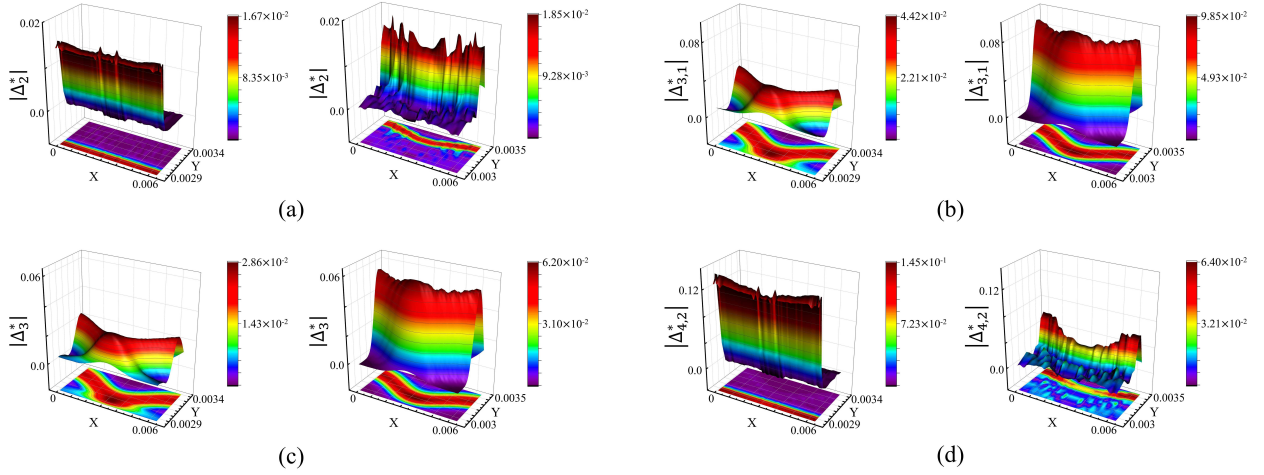}
\caption{Different orders of the nonequilibrium intensity at the substance interface before and after the shock wave passes through the interface for the first time: (a) $\left |\bm{\Delta} _2^{ {\rm{*}}}\right |$, (b) $\left |\bm{\Delta} _{3,1}^{ {\rm{*}}}\right |$, (c) $\left |\bm{\Delta} _{3}^{ {\rm{*}}}\right |$, and (d) $\left |\bm{\Delta} _{4,2}^{ {\rm{*}}}\right |$; the left and right images of each picture represent before and after the shock wave passes through the interface, respectively.}
\label{Fig:RMI2}
\end{figure*}

Next, we analyze the nonequilibrium kinetics effect of single-mode RMI and reshock processes based on the DBM simulation results. For the RMI that occurs when the shock wave passes through the substance interface for the first time, Fig. \ref{Fig:RMI2} shows the different orders of the nonequilibrium intensity at the substance interface, and (a), (b), (c) and (d) correspond to $\left |\bm{\Delta} _2^{ {\rm{*}}}\right |$, $\left |\bm{\Delta} _{3,1}^{ {\rm{*}}}\right |$, $\left |\bm{\Delta} _{3}^{ {\rm{*}}}\right |$ and $\left |\bm{\Delta} _{4,2}^{ {\rm{*}}}\right |$ before (left) and after (right) the shock front passes through the interface, respectively. According to the mathematical derivation, $\bm{\Delta} _{2}^{\rm{*}}$ (NOMF) and $\bm{\Delta} _{3,1}^{\rm{*}}$ (NOEF) correspond to the viscous stress and heat flux in fluid dynamics, respectively, $\bm{\Delta} _{3}^{\rm{*}}$ is equivalent to the flux of $\bm{\Delta} _{2}^{\rm{*}}$, and $\bm{\Delta} _{4,2}^{\rm{*}}$ is equivalent to the flux of $\bm{\Delta} _{3,1}^{\rm{*}}$. It is essential to stress that although corresponding terms for $\bm{\Delta} _{3}^{\rm{*}}$ and $\bm{\Delta} _{4,2}^{\rm{*}}$ are not found in the macroscopic hydrodynamic equations, these are physical quantities derived from the Boltzmann equation, which has a more fundamental theory than the Euler or Navier-Stokes equations, and they cannot be obtained by the Euler or Navier-Stokes equations. In Fig. \ref{Fig:RMI2} (a), we find that the shock wave causes a very strong nonequilibrium effect, while the nonequilibrium effect at the interface is inconspicuous. However, when the shock wave passes through the substance interface, the shock front is deformed due to the presence of the substance interface, and a nonequilibrium effect exists at the substance interface due to the shock effects. As shown in Fig. \ref{Fig:RMI2} (b), the nonequilibrium effect $\left |\bm{\Delta} _{3,1}^{ {\rm{*}}}\right |$ (which corresponds to the amplitude of the heat flux) is generated by two factors: the difference in temperature between the two substances and the shock wave. This characteristic is different from $\left |\bm{\Delta} _2^{ {\rm{*}}}\right |$. Specifically, nonequilibrium effects due to the strong interruption caused by the shock front and nonequilibrium effects at the substance interface due to the heat conduction between the two substances can be observed when the shock front has not yet reached the substance interface. When the shock front interacts with the substance interface, $\left |\bm{\Delta} _{3,1}^{ {\rm{*}}}\right |$ significantly increases, indicating that the shock wave causes substances at the interface to deviate from the equilibrium state more severely and that heat conduction at the substance interface increases. In Fig. \ref{Fig:RMI2} (c), $\left |\bm{\Delta} _{3}^{ {\rm{*}}}\right |$(corresponding to the flux of $\left |\bm{\Delta} _2^{ {\rm{*}}}\right |$) exhibits similar images to $\left |\bm{\Delta} _{3,1}^{ {\rm{*}}}\right |$. In Fig. \ref{Fig:RMI2} (d), the shock wave exhibits nonequilibrium effects before interacting with the substance interface, which is similar to $\left |\bm{\Delta} _2^{ {\rm{*}}}\right |$. However, $\left |\bm{\Delta} _{4,2}^{ {\rm{*}}}\right |$ decreases when the shock front impacts the substance interface, and complex nonequilibrium kinetic effects are also present at the substance interface due to the transmission and reflection of the shock wave. Greatly different from the case of normal shock on an unperturbed plane interface between two uniform media, which is well described by the Rankine-Hugoniot relations and where all TNE quantities quickly recover to zero behind the shock front, in the RMI case, the TNE quantities show complex but inspiring kinetic effects in the shock process and behind the shock front.

\begin{figure*}[ht]
\centering
\includegraphics[height=7cm]{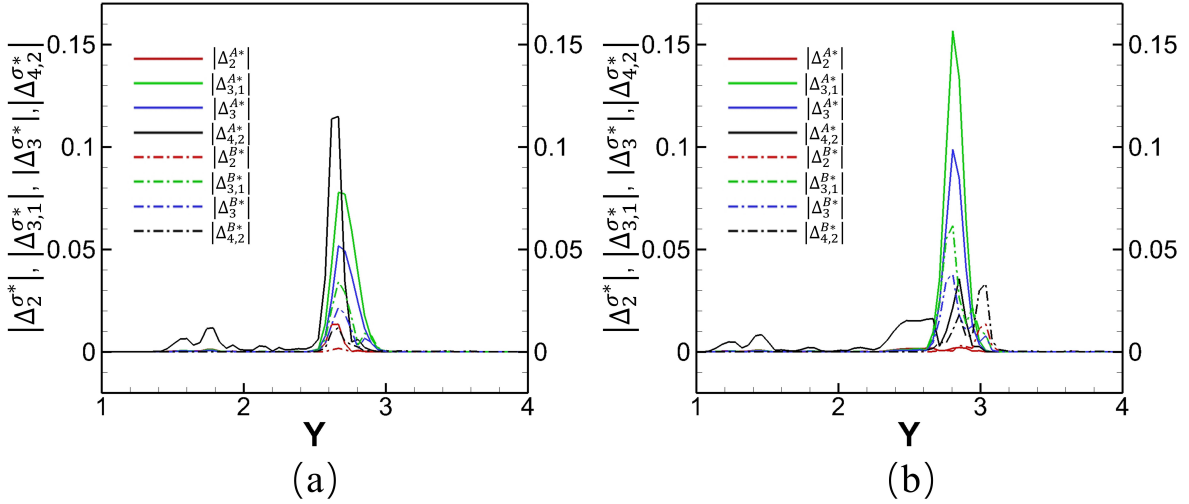}
\caption{Nonequilibrium quantity distribution of substances A and B along the $y$-axis through the bubble before (a) and after (b) the shock.}
\label{Fig:RMI3}
\end{figure*}

\begin{figure*}[ht]
\centering
\includegraphics[height=7cm]{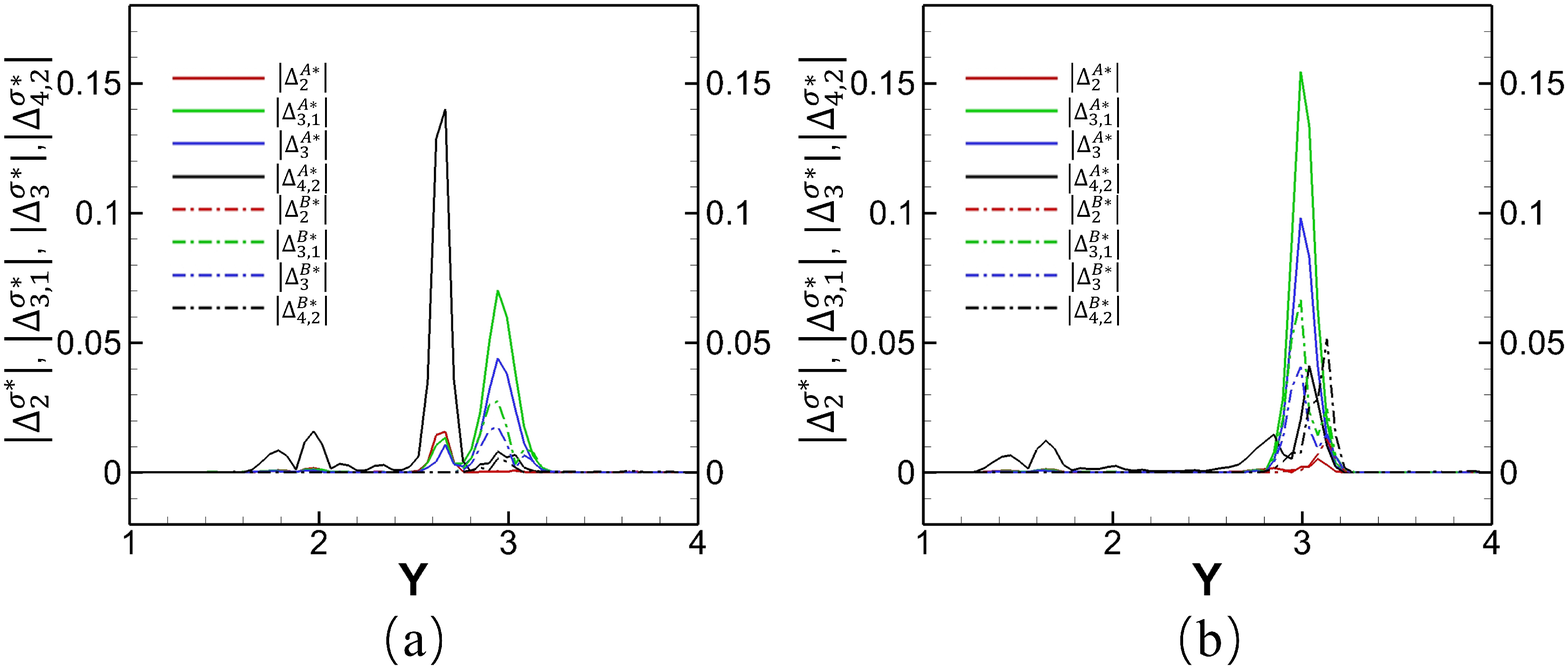}
\caption{Nonequilibrium quantity distribution of substances A and B along the $y$-axis through the spike before (a) and after (b) the shock.}
\label{Fig:RMI4}
\end{figure*}

For further analysis, we display the one-dimensional different orders of the nonequilibrium quantity intensities of substance A and substance B before and after the occurrence of RMI. Specifically, the distributions of nonequilibrium quantities along the $y$-axis through the RMI spike and the bubble are shown. The nonequilibrium quantity distributions of substances A and B along the $y$-axis through the bubble before and after the shock are depicted in Fig. \ref{Fig:RMI3}, and Fig. \ref{Fig:RMI4} presents those through the spike. Figures \ref{Fig:RMI3} and \ref{Fig:RMI4} illustrate that each order of the nonequilibrium intensity of substance A is typically greater than that of B. Specifically, (1) $\left |\bm{\Delta} _2^{A {\rm{*}}}\right |$ decreases relative to its value before the shock, while $\left |\bm{\Delta} _2^{B {\rm{*}}}\right |$ becomes greater than its value before the shock. $\left |\bm{\Delta} _{4,2}^{A {\rm{*}}}\right |$ and $\left |\bm{\Delta} _{4,2}^{B {\rm{*}}}\right |$ are similar to $\left |\bm{\Delta} _2^{A {\rm{*}}}\right |$ and $\left |\bm{\Delta} _2^{B {\rm{*}}}\right |$. (2) $\left |\bm{\Delta} _{3,1}^{A {\rm{*}}}\right |$ and $\left |\bm{\Delta} _{3,1}^{B {\rm{*}}}\right |$ become greater than their values before the shock. $\left |\bm{\Delta} _{3}^{A {\rm{*}}}\right |$ and $\left |\bm{\Delta} _{3}^{B {\rm{*}}}\right |$ are similar to $\left |\bm{\Delta} _{3,1}^{A {\rm{*}}}\right |$ and $\left |\bm{\Delta} _{3,1}^{B {\rm{*}}}\right |$. (3) In general, the locations of the peaks of different orders of the nonequilibrium quantity intensities for substances A and B do not overlap. The above conclusions indicate that there are different effects acting on substances A and B even for the same process. There are also differences in the changes in different nonequilibrium quantities for the same substance.

\begin{figure*}[ht]
\centering
\includegraphics[height=9cm]{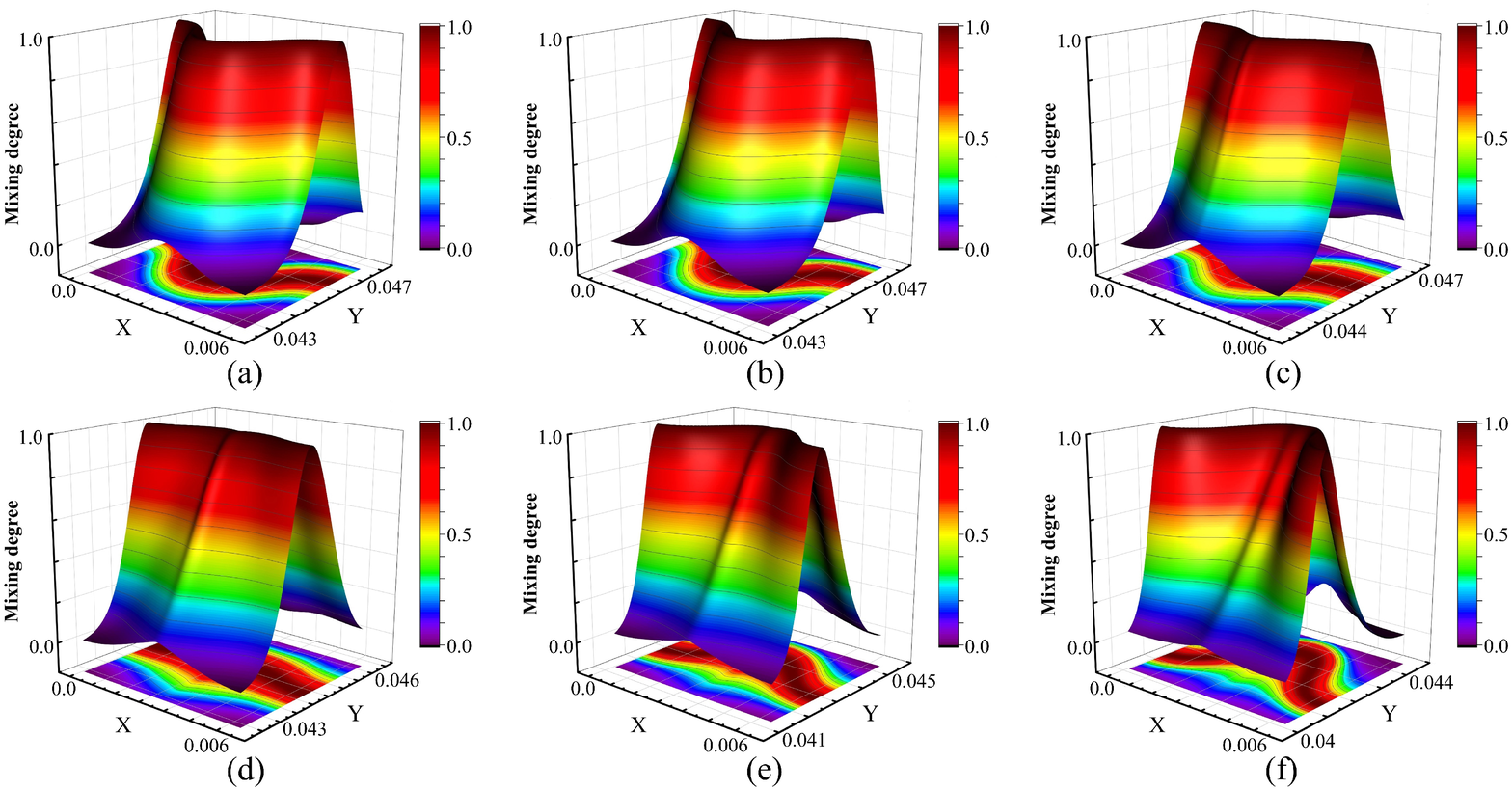}
\caption{Mixing degree versus time in the mixing zone during reshock: (a) before the reshock wave reaches the mixing zone at $t=0.246$; (b) to (f): the reshock wave interacts with the substance interface at $t=0.252, 0.258, 0.264, 0.270$ and $0.276$.}
\label{Fig:RS1}
\end{figure*}

As the shock front continues to propagate along the positive $y$-axis, it interacts with the upper boundary at some point, generating a reflected wave propagating in the negative $y$-axis direction, and the reflected wave interacts with the substance interface again, i.e., reshock. For reshock, we define the mixing degree $M_{ix}$:
\begin{equation}
M_{ix}= 4{F_\mathrm{A}}{F_\mathrm{B}},
\end{equation}
where ${F_\mathrm{A}}$ and ${F_\mathrm{B}}$ are the mass fractions of substances A and B, respectively. Therefore, the range of $M_{ix}$ is $0 \le M_{ix} \le 1$. $M_{ix}=1$  means complete mixing, and $M_{ix}=0$  means no mixing. The change in the mixing degree for the two substances in the mixing zone before and after the reshock is depicted in Fig. \ref{Fig:RS1}. The mixing degree at the substance interface is 1, which represents the most fully mixed positions, and it decreases from the interface outward. From Figs. \ref{Fig:RS1} (b) to (f), the shock front starts to interact with the interface, it can be seen that the range of the mixing zone is initially compressed and then expanded, and the RMI interface reverses due to reshock.

\begin{figure*}[ht]
\centering
\includegraphics[height=9cm]{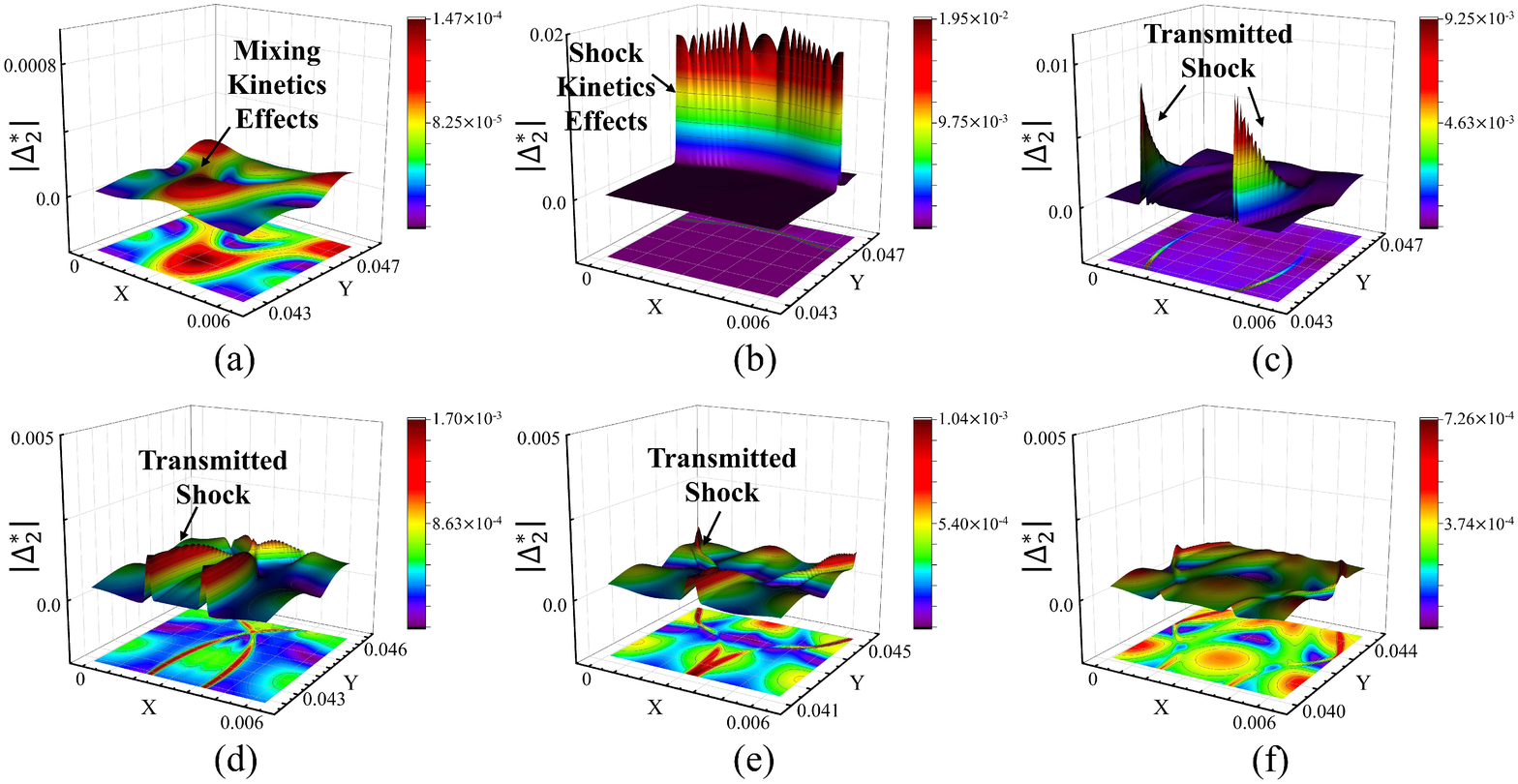}
\caption{$\left |\bm{\Delta} _2^{ {\rm{*}}}\right |$ versus time in the mixing zone during reshock; (a) to (f) in this figure correspond to (a) to (f) in Fig. \ref{Fig:RS1}, respectively.}
\label{Fig:RSD2}
\end{figure*}

\begin{figure*}[ht]
\centering
\includegraphics[height=9cm]{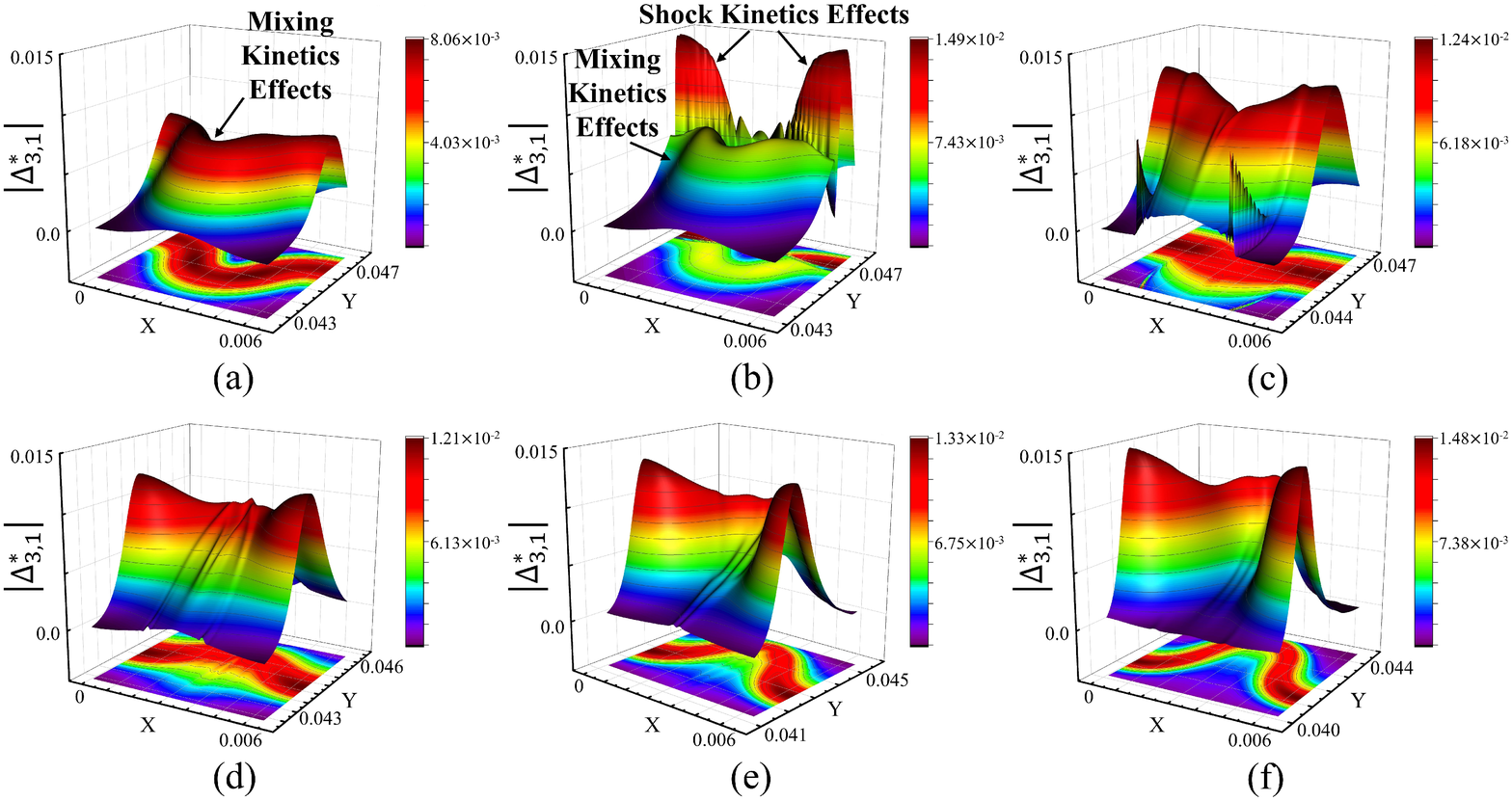}
\caption{$\left |\bm{\Delta} _{3,1}^{ {\rm{*}}}\right |$ versus time in the mixing zone during reshock; (a) to (f) in this figure correspond to (a) to (f) in Fig. \ref{Fig:RS1}, respectively.}
\label{Fig:RSD31}
\end{figure*}

\begin{figure*}[ht]
\centering
\includegraphics[height=9cm]{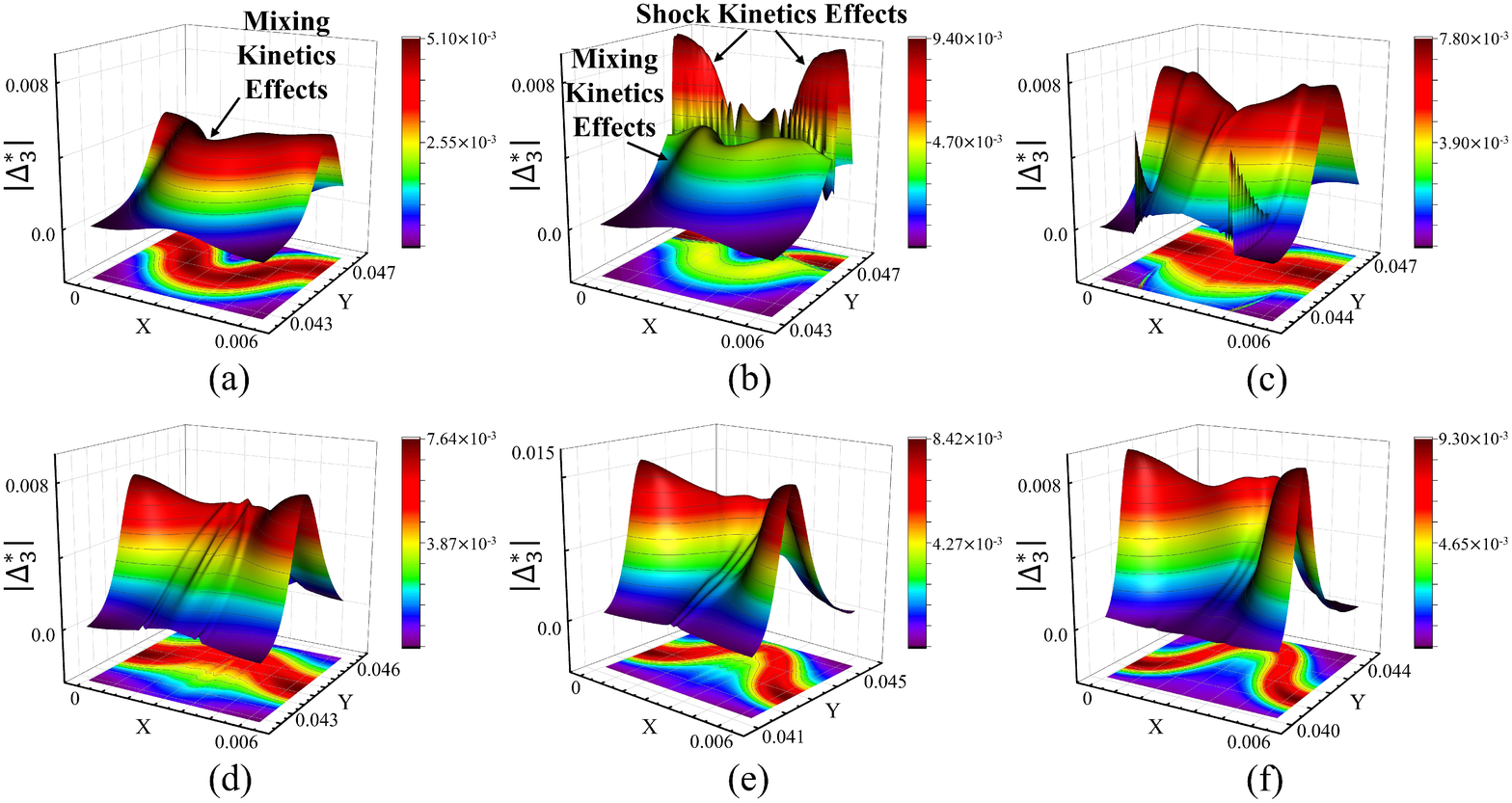}
\caption{$\left |\bm{\Delta} _{3}^{ {\rm{*}}}\right |$ versus time in the mixing zone during reshock; (a) to (f) in this figure correspond to (a) to (f) in Fig. \ref{Fig:RS1}, respectively.}
\label{Fig:RSD3}
\end{figure*}

\begin{figure*}[ht]
\centering
\includegraphics[height=9cm]{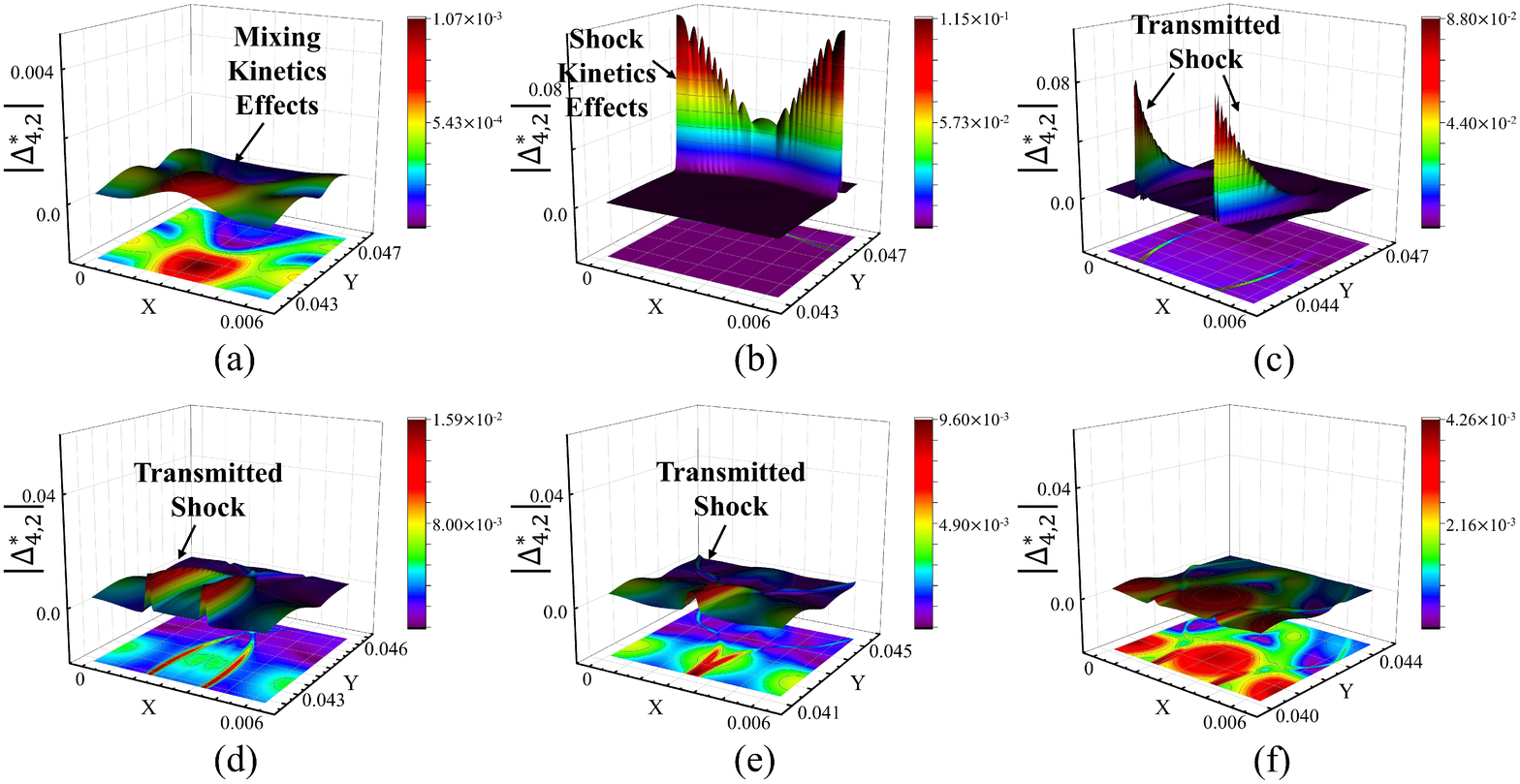}
\caption{$\left |\bm{\Delta} _{4,2}^{ {\rm{*}}}\right |$ versus time in the mixing zone during reshock; (a) to (f) in this figure correspond to (a) to (f) in Fig. \ref{Fig:RS1}, respectively.}
\label{Fig:RSD42}
\end{figure*}

In the reshock process, we calculate the evolution of the different orders of the nonequilibrium intensity $\left |\bm{\Delta} _2^{ {\rm{*}}}\right |$, $\left |\bm{\Delta} _{3,1}^{ {\rm{*}}}\right |$, $\left |\bm{\Delta} _{3}^{ {\rm{*}}}\right |$, and $\left |\bm{\Delta} _{4,2}^{ {\rm{*}}}\right |$ (see Fig. \ref{Fig:RSD2} to Fig. \ref{Fig:RSD42}), where (a) to (f) in these figures correspond to (a) to (f) in Fig. \ref{Fig:RS1}, respectively. For $\left |\bm{\Delta} _2^{ {\rm{*}}}\right |$, Fig. \ref{Fig:RSD2}(a) displays that before the shock front reaches the mixing zone, there are nonequilibrium effects in the mixing zone because the RMI is developing. When the shock front reaches the mixing zone, we can observe in Fig. \ref{Fig:RSD2}(b) that an interrupted interface is caused by the strong nonequilibrium effects of the shock wave, whose values are much greater than the nonequilibrium intensity in the mixing zone. In Fig. \ref{Fig:RSD2}(c), the reshock wave reaches the interface, is reflected and transmits within the mixing zone, and the shock front deforms. For Figs. \ref{Fig:RSD2} (d) to (f), since the reshock wave has already passed through the mixing zone, what propagates in the mixing zone at this time is the transmitted shock generated by the interaction between the shock front and substance interface. The nonequilibrium situation in the mixing zone is complicated, but the intensity is generally decreasing.

Figure \ref{Fig:RSD31} shows the $\left |\bm{\Delta} _{3,1}^{ {\rm{*}}}\right |$ in the mixing zone during the reshock. Figure \ref{Fig:RSD31}(a) indicates that when the reshock wave has not yet reached the substance mixing zone, $\left |\bm{\Delta} _{3,1}^{ {\rm{*}}}\right |$ exists at the substance interface due to the temperature difference between substances A and B, and its shape and scale are essentially identical to those of the substance mixing layer. The shock front has started to interact with the mixing interface in Fig. \ref{Fig:RSD31}(b), and compared to the nonequilibrium intensity generated by the shock wave only, $\left |\bm{\Delta} _{3,1}^{ {\rm{*}}}\right |$ increases under the combined effects of the nonequilibrium intensity already present at the interface and the nonequilibrium intensity generated by the shock wave. On the other hand, we can conclude from this diagram that the magnitude of $\left |\bm{\Delta} _{3,1}^{ {\rm{*}}}\right |$ generated by the shock wave is on the same order as that of $\left |\bm{\Delta} _{3,1}^{ {\rm{*}}}\right |$ generated by substance mixing, and all of the above conclusions about $\left |\bm{\Delta} _{3,1}^{ {\rm{*}}}\right |$ differ from those for $\left |\bm{\Delta} _2^{ {\rm{*}}}\right |$. The $\left |\bm{\Delta} _{3,1}^{ {\rm{*}}}\right |$ generated by the mixing interacts continuously with the $\left |\bm{\Delta} _{3,1}^{ {\rm{*}}}\right |$ generated by the shock wave in Fig. \ref{Fig:RSD31}(c), and the $\left |\bm{\Delta} _{3,1}^{ {\rm{*}}}\right |$ intensity at this point in time is greater than that of $\left |\bm{\Delta} _{3,1}^{ {\rm{*}}}\right |$ in Fig. \ref{Fig:RSD31}(a), indicating that the shock wave intensifies the deviation of the fluid system from equilibrium in the mixing zone. In Figs. \ref{Fig:RSD31}(d) to (f), the reshock wave has traversed the mixing zone, and the substance interface is reversed. $\left |\bm{\Delta} _{3,1}^{ {\rm{*}}}\right |$ reverses concurrently with the substance interface. Notably, $\left |\bm{\Delta} _{3,1}^{ {\rm{*}}}\right |$ increases as the interface reverses because the reshock intensifies the mixing of the substances, thereby increasing the nonequilibrium degree.

Figures \ref{Fig:RSD3} and \ref{Fig:RSD42} show the evolution of $\left |\bm{\Delta} _{3}^{ {\rm{*}}}\right |$ and $\left |\bm{\Delta} _{4,2}^{ {\rm{*}}}\right |$ in the mixing zone during the reshock. It has been discovered that the diagrams of $\left |\bm{\Delta} _{4,2}^{ {\rm{*}}}\right |$ and $\left |\bm{\Delta} _2^{ {\rm{*}}}\right |$, $\left |\bm{\Delta} _{3}^{ {\rm{*}}}\right |$ and $\left |\bm{\Delta} _{3,1}^{ {\rm{*}}}\right |$ are comparable, respectively, but they are not identical. In addition, the distribution of each order of the nonequilibrium quantities in the reshock demonstrates that the nonequilibrium quantity diagrams exhibit a pattern similar to that obtained by the schlieren method, which is commonly used in shock wave experiments, and display flow field details that cannot be obtained by direct observation of the flow field. This demonstrates that, similar to the schlieren method, the nonequilibrium quantities help grasp the flow details within the flow fields, which is one of the advantages of the DBM.
The above results depict the global nonequilibrium effect of the mixing zone during the reshock process. Due to the implementation of the two-fluid DBM in this study, each substance has its independent distribution function $f$ and Boltzmann evolution equation, allowing information regarding the flow of each substance to be separately provided, which is unavailable using the single-fluid simulation method. To analyze the detailed behavior of each substance during the reshock process, here, we provide the nonequilibrium intensity distributions of substances A and B in the mixing zone.

\begin{figure*}
\centering
\includegraphics[height=11cm]{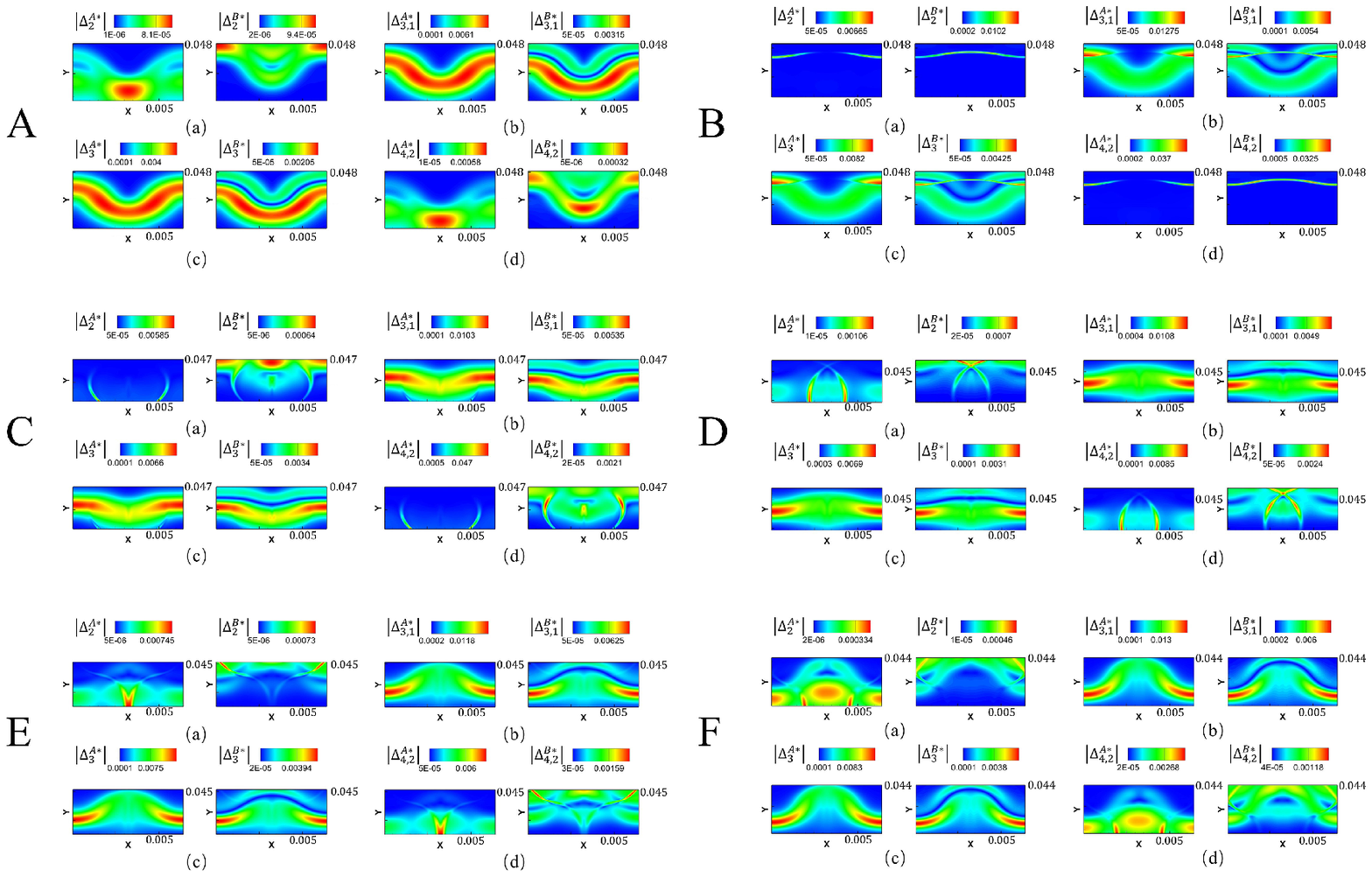}
\caption{Different orders of the nonequilibrium intensity of substances A and B in the reshock process. (A) to (F) correspond to the six moments referred to in Fig.\ref{Fig:RS1}.}
\label{Fig:RS3}
\end{figure*}

Figure \ref{Fig:RS3} shows the various orders of the nonequilibrium intensity of each substance in the reshock process. It comprises graphs (A) to (F), which correspond to the six moments in Fig. \ref{Fig:RS1}, similar to the previous figures. In addition, (A) to (F) also contain (a), (b), (c) and (d), which represent $\left |\bm{\Delta} _2^{ {\rm{*}}}\right |$, $\left |\bm{\Delta} _{3,1}^{ {\rm{*}}}\right |$, $\left |\bm{\Delta} _{3}^{ {\rm{*}}}\right |$ and $\left |\bm{\Delta} _{4,2}^{ {\rm{*}}}\right |$, respectively. Overall, we can observe that there are overlaps between the nonequilibrium intensity diagrams of substances A and B due to the diffusion effect between the two substances, as shown in Fig. \ref{Fig:RS3}. For the spike and bubble structures in Fig. \ref{Fig:RS3} (A), which are already in a nonlinearly stable development stage, the nonequilibrium effects generated by substances A and B at the spike and bubble are not consistent. However, in general, the nonequilibrium intensity $\left |\bm{\Delta} _2^{ {\rm{*}}}\right |$ is greater within the substance that is ``forced'' to change its flow state, as shown in the $\left |\bm{\Delta} _2^{ {\rm{*}}}\right |$ part of Fig. \ref{Fig:RS3} (A) (for example, since the spike develops in substance A, A is here called the substance that is ``forced'' to change its flow state, and the opposite for the bubble). It can be seen from Fig. \ref{Fig:RS3} (B) that although the existence of temperature differences between two substances leads to the generation of a heat flux at the macroscopic level if the two substances are viewed separately, there is a difference in the heat flux generated by each of the two substances for the one process due to their different characteristics. The macroscopic manifestation of the flow system behavior is determined by the combined response of both substances. In Fig. \ref{Fig:RS3} (B), the reshock wave has reached the mixing zone and started to interact with the interfacial material, and the nonequilibrium effect due to the presence of the shock wave is captured in the nonequilibrium intensity images of both the A and B substances, whereas the nonequilibrium effect due to the RMI evolution within the mixing interface is not obvious in the $\left |\bm{\Delta} _2^{ {\rm{*}}}\right |$ and $\left |\bm{\Delta} _{4,2}^{ {\rm{*}}}\right |$ profiles because it is too small in comparison to the nonequilibrium effect produced by the shock front. This conclusion is identical to that for the first shock. In addition to leading to discontinuity of macroscopic physical quantities, the shock wave also leads to discontinuity of nonequilibrium quantities, as well as expansion of the nonequilibrium region and intensification of the nonequilibrium effect, as shown in the $\left |\bm{\Delta} _{3}^{ {\rm{*}}}\right |$ and $\left |\bm{\Delta} _{3,1}^{ {\rm{*}}}\right |$ profiles. In Fig. \ref{Fig:RS3} (C), the propagation of the shock front in the mixing zone is captured. Specifically, the $\left |\bm{\Delta} _2^{ {\rm{*}}}\right |$ and $\left |\bm{\Delta} _{4,2}^{ {\rm{*}}}\right |$ profiles depict the propagation of the shock front in the mixing zone, and in the upper portion of the $\left |\bm{\Delta} _2^{ {\rm{*}}}\right |$ and $\left |\bm{\Delta} _{4,2}^{ {\rm{*}}}\right |$ profiles of substance B, three regions exhibit a high degree of nonequilibrium intensity. The distributions of $\left |\bm{\Delta} _2^{ {\rm{*}}}\right |$ and $\left |\bm{\Delta} _{4,2}^{ {\rm{*}}}\right |$ reveal that the spike is reversing at this point as a result of the shock. The above is also illustrated in the $\left |\bm{\Delta} _{3,1}^{ {\rm{*}}}\right |$ and $\left |\bm{\Delta} _{3}^{ {\rm{*}}}\right |$ distributions. Despite the fact that this conclusion can be reached from both profiles, the phenomenon is not expressed in the same way. In Figs. \ref{Fig:RS3} (D) to (F), the reshock wave has already passed through the mixing zone. However, due to the interaction between the shock front and substance interface, the transmitted shock waves propagate within the mixing zone. In addition to the nonequilibrium effect generated by the shock wave, the nonequilibrium effect generated by the interface reversal and mixing at the substance interface is also captured, and the two kinds of nonequilibrium effects interact with each other.

\begin{figure*}[ht]
\centering
\includegraphics[height=7cm]{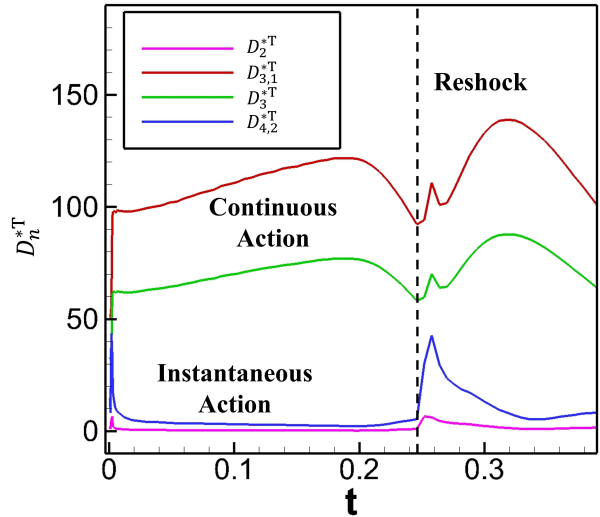}
\caption{Variation of different orders of the nonequilibrium intensities in the mixing zone with time $t$.}
\label{Fig:RS4}
\end{figure*}

Figure \ref{Fig:RS4} shows the temporal evolution of the nonequilibrium intensity vector
$\mathbf{D}^{\rm{neq}}$ which shows
the nonequilibrium intensities from four different perspectives in the mixing zone. Near the initial moment, there is a jump in the nonequilibrium intensities caused by the interaction of the first shock, and then, the RMI process starts to develop. As the material interface develops, $D^{\rm{*T}}_{3,1}$ and $D^{\rm{*T}}_{3}$ slowly increase and then decrease, while $D^{\rm{*T}}_{2}$ and $D^{\rm{*T}}_{4,2}$ remain essentially unchanged after a rapid reduction with the shock front passing through the mixing zone. In summary, the shock wave has an ``\emph{instantaneous action}'' for $D^{\rm{*T}}_{2}$ (the intensity of the viscous stress) and $D^{\rm{*T}}_{4,2}$ but a ``\emph{continuous action}'' for $D^{\rm{*T}}_{3,1}$ (the intensity of the heat flux) and $D^{\rm{*T}}_{3}$ in the mixing zone of the RMI. This means that the shock action greatly enhances the heat flux of the fluid, and this enhancement is always present even when the shock front passes through and does not interact with the mixing interface. However, shock waves have the opposite effects on the viscous stress. At $t$=0.246, reshock occurs, and the nonequilibrium intensities increase once more. For $D^{\rm{*T}}_{3,1}$ and $D^{\rm{*T}}_{3}$, the increments of the nonequilibrium intensity produced by reshock are less than the value at the first shock, whereas the increments of $D^{\rm{*T}}_{2}$ and $D^{\rm{*T}}_{4,2}$ are roughly equivalent compared with the first shock. The reasons are that the reshock process reverses the velocity of the fluid at the interface and the range of the mixing zone is larger than that for the first shock, so although the intensity of the reshock wave is weaker than that of the shock at the initial time, the increase in the nonequilibrium intensity at reshock is equivalent to that of the RMI process at the initial time for $D^{\rm{*T}}_{2}$ and $D^{\rm{*T}}_{4,2}$. After reshock, $D^{\rm{*T}}_{3,1}$ and $D^{\rm{*T}}_{3}$ once more exhibit an increase followed by a decrease, and $D^{\rm{*T}}_{2}$ and $D^{\rm{*T}}_{4,2}$ remain essentially constant after a decrease.

\begin{figure*}[ht]
\centering
\includegraphics[height=7cm]{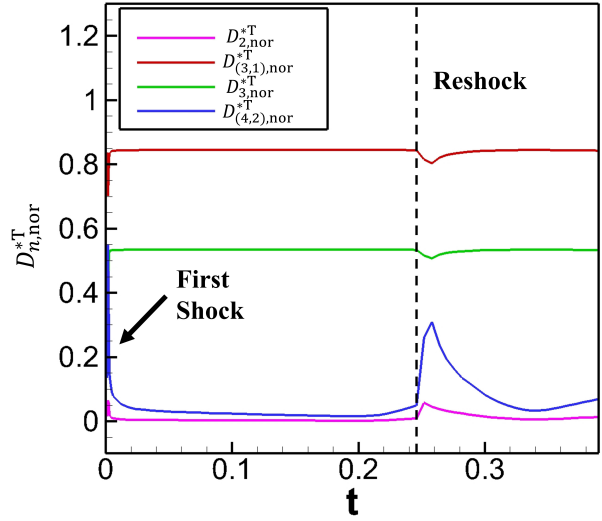}
\caption{Normalized nonequilibrium intensity quantities in the mixing zone with time $t$.
\label{Fig:RS6}}
\end{figure*}

Figure \ref{Fig:RS6}
shows the temporal evolution of the nonequilibrium intensity vector
$\mathbf{D}^{\rm{neq}}_{\rm{nor}}$ whose components are non-dimensionalized and normalized nonequilibrium intensity quantities  $D^{\rm{*T}}_{n,\rm{nor}}$ and which shows
the nonequilibrium intensities from four different perspectives in the mixing zone.
It is clear that Figs. \ref{Fig:RS6} and \ref{Fig:RS4} show the same kinetic behaviors in two different perspective.
 The following results can be obtained: $D^{\rm{*T}}_{(3,1),\rm{nor}}$ has the greatest contribution to the total nonequilibrium intensity. $D^{\rm{*T}}_{2,\rm{nor}}$, $D^{\rm{*T}}_{(3,1),\rm{nor}} $, and $D^{\rm{*T}}_{3,\rm{nor}}$ account for a nearly constant proportion of ${D_{\mathrm{sum}}}$ throughout the entire process except the first shock and reshock, but $D^{\rm{*T}}_{(4,2),\rm{nor}}$ experience an abrupt and significant change when the shock wave interacts with the substance interface.
 The properties of the nonequilibrium intensity vector can be utilized as the sign of the interaction between the shock waves and substance interface.

The definition of the molecular mixing fraction/ratio\upcite{youngs1991three,youngs1994numerical} is:
\begin{equation}
R = \frac{{\overline {{F_\mathrm{A}} \cdot {F_\mathrm{B}}} }}{{\overline {{F_\mathrm{A}}}  \cdot \overline {{F_\mathrm{B}}} }}.
\end{equation}

\begin{figure*}[ht]
\centering
\includegraphics[height=7cm]{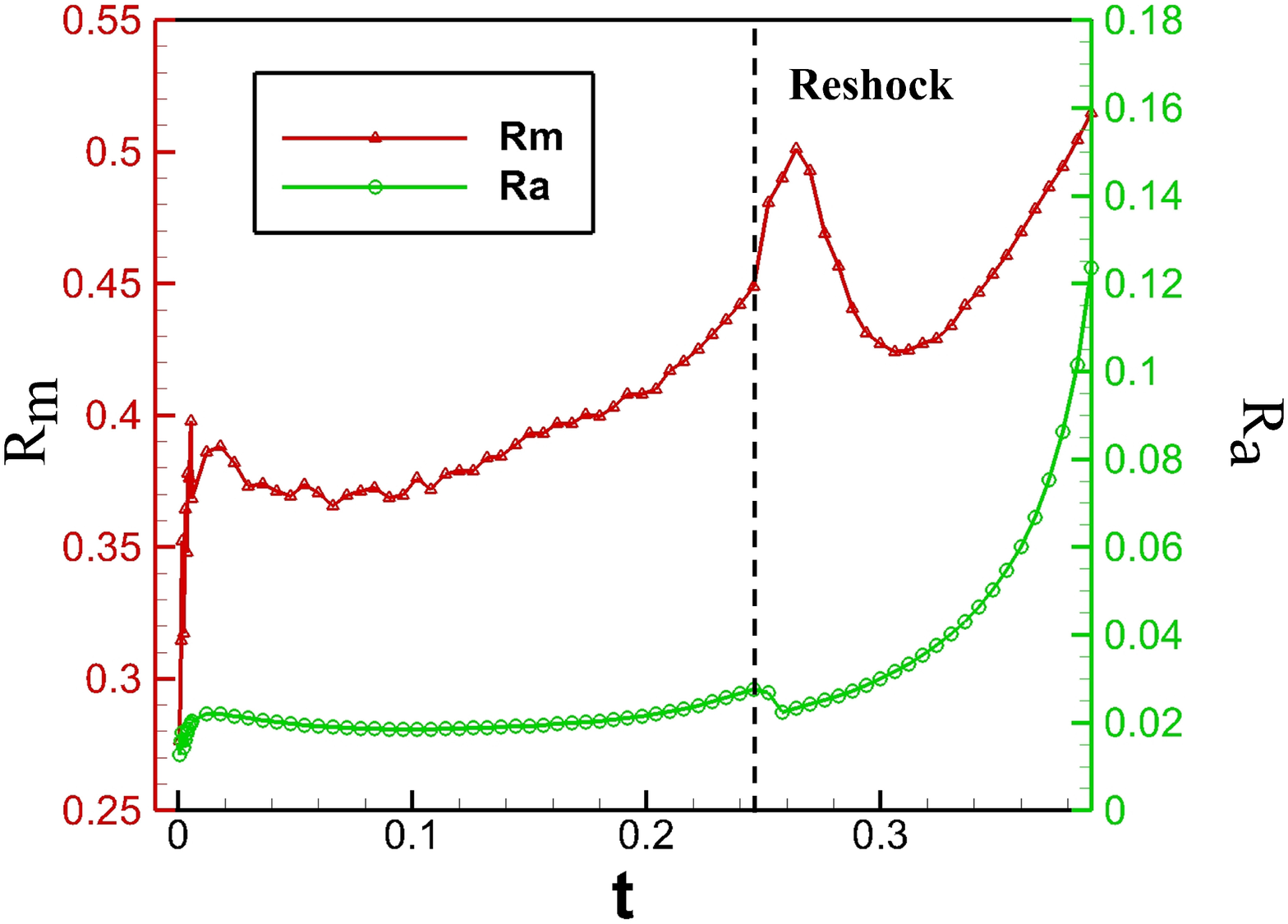}
\caption{Molecular mixing fraction profiles versus time $t$. ${R_m}$: mixing zone; ${R_a}$: whole flow field.}
\label{Fig:RS5}
\end{figure*}
The molecular mixing fractions of the mixing zone and the whole flow field are denoted by ${R_m}$ and ${R_a}$, respectively. Figure \ref{Fig:RS5} shows the molecular mixing fraction ${R_m}$ and ${R_a}$ profiles versus time $t$. Even for the same process, different study scopes result in distinct variations in the molecular mixing fraction, as shown in Fig. \ref{Fig:RS5}. For ${R_m}$, RMI begins to occur when the initial shock front impacts the interface, and ${R_m}$ significantly increases. Then, as the shock front passes through the mixing zone and RMI continues to develop, ${R_m}$ slowly increases as the fluid instability develops. Reshock occurs at $t$=0.246, and at the stage when the reshock wave interacts with the substance interface, both the decrease in the width of the mixing zone due to the shock front impacting the mixing zone and the promotion of the mixing of the substances caused by the reshock result in a rapid increase in ${R_m}$. Then, the mixing zone starts to expand as the shock front passes through the mixing zone, which corresponds to the inclusion of unmixed fluid into the mixing zone, resulting in a decrease in the degree of mixing in the mixing zone. However, as the RMI continues to develop, ${R_m}$ increases, and the rate is greater than that in the initial shock, which indicates that reshock can promote the mixing of substances in the flow field.

For ${R_a}$, initially, ${R_a}$ increases when RMI begins to develop after the shock front impacts the interface. Subsequently, ${R_a}$ gradually increases as the RMI interface evolves. When $t$ = 0.246, reshock occurs, and ${R_a}$ declines, which is contrary to the variation in ${R_m}$ at this time. This occurs because although the degree of mixing increases, the area of the mixing zone is reduced, leading to a decrease in ${R_a}$. However, when the shock front passes through the mixing zone, ${R_a}$ begins to increase at a much faster rate than the first shock. The above shows that reshock promotes the flow field substance mixing, but the degree of mixing does not monotonically increase.
\subsection{Entropy production rate}

Entropy is one of the key factors affecting the success of the final ignition in ICF implosion, and the entropy change is a crucial issue in ICF research. Therefore, the study of entropy is of great importance to engineering applications. The above is an analysis of the nonequilibrium effects present in the RMI and reshock processes, and the entropy production rates can be another set of TNE quantities. To gain a profound comprehension of RMI and reshock processes, we discuss the entropy production rates below.
\begin{figure*}[ht]
\centering
\includegraphics[height=5cm]{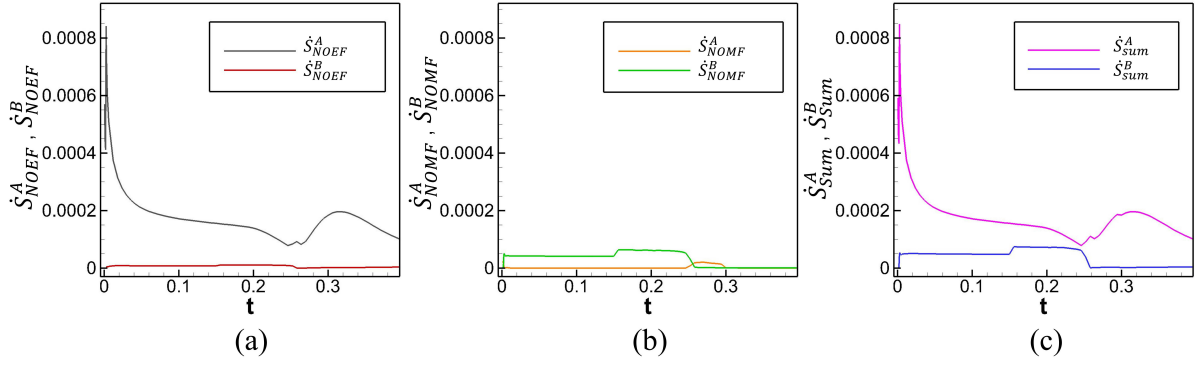}
\caption{Entropy production rate of substances A and B in the whole field. (a) ${\dot{S} _{\mathrm{NOEF}} }$ of substances A and B; (b) ${\dot{S} _{\mathrm{NOMF}} }$ of substances A and B; (c) entropy production rate ${\dot{S} _{\mathrm{sum}} }$ of substances A and B.}
\label{Fig:S1}
\end{figure*}

Figure \ref{Fig:S1} depicts the variation in the entropy production rates of substances A and B in the whole flow field over time $t$. Figure \ref{Fig:S1} (a) shows a significant increase in $\dot{S} _{\mathrm{NOEF}}^{\mathrm{A}} $ at the initial moment due to the shock effect, followed by a gradual decrease as RMI develops. There is a small increase in $\dot{S} _{\mathrm{NOEF}}^{\mathrm{A}} $ when reshock occurs, and its increment is less than the increment in the entropy production rate caused by the shock initially impacting the interface, because the intensity of reshock is weaker than that of the initial shock. $\dot{S} _{\mathrm{NOEF}}^{\mathrm{A}} $ then significantly increases because the reshock promotes the mixing of substances, resulting in an enhancement of $\dot{S} _{\mathrm{NOEF}}^{\mathrm{A}} $. Eventually, $\dot{S} _{\mathrm{NOEF}}^{\mathrm{A}} $ gradually decreases as the mixing process stabilizes. For $\dot{S} _{\mathrm{NOEF}}^{\mathrm{B}} $ in Fig. \ref{Fig:S1}(a), the value initially increases when the shock front enters substance B after interacting with the substance interface, and afterward, $\dot{S} _{\mathrm{NOEF}}^{\mathrm{B}} $ remains essentially constant until the shock front re-enters substance A after reshock; thus, $\dot{S} _{\mathrm{NOEF}}^{\mathrm{B}} $ tends to 0. Regarding $\dot{S} _{\mathrm{NOMF}}^{\mathrm{B}} $, as shown in Fig. \ref{Fig:S1}(b), initially, the shock front enters substance B from A after interacting with the substance interface. Consequently, $\dot{S} _{\mathrm{NOMF}}^{\mathrm{B}} $ increases during the initial stage and then remains essentially constant. At $t=0.156$, the shock front moves to the upper wall surface and is reflected in the negative $y$-axis direction, thereby causing $\dot{S} _{\mathrm{NOMF}}^{\mathrm{B}} $ to increase once more. When reshock occurs, the shock front from substance B enters A again, $\dot{S} _{\mathrm{NOMF}}^{\mathrm{B}} $ tends to 0, and $\dot{S} _{\mathrm{NOMF}}^{\mathrm{A}} $ increases; however, as the shock front moves out of the flow field, $\dot{S} _{\mathrm{NOMF}}^{\mathrm{A}} $ tends to 0 at the end. For the entropy production rate ${\dot{S} _{\mathrm{sum}} }$ in Fig. \ref{Fig:S1}(c), the contribution of $\dot{S} _{\mathrm{NOEF}}^{\mathrm{A}} $ causes the entropy production rate of substance A $\dot{S} _{\mathrm{sum}}^{\mathrm{A}} $ to be greater than that of substance B $\dot{S} _{\mathrm{sum}}^{\mathrm{B}} $.
\begin{figure*}[ht]
\centering
\includegraphics[height=5cm]{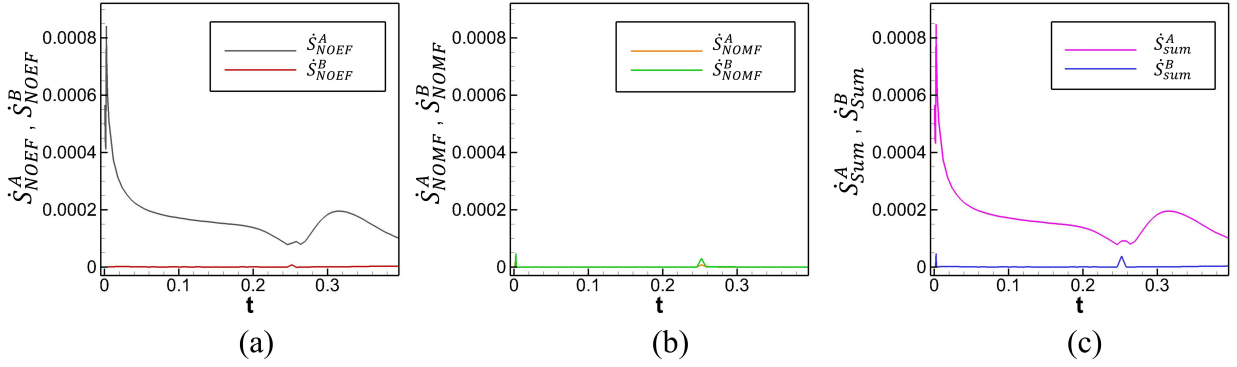}
\caption{Entropy production rates of substances A and B in the mixing zone. (a) ${\dot{S} _{\mathrm{NOEF}} }$ of substances A and B; (b) ${\dot{S} _{\mathrm{NOMF}} }$ of substances A and B; (c) entropy production rate ${\dot{S} _{\mathrm{sum}} }$ of substances A and B.}
\label{Fig:S2}
\end{figure*}

Similarly, the variation in the entropy production rates of substances A and B in the mixing zone with time is shown in Fig. \ref{Fig:S2}. Compared to Fig. \ref{Fig:S1}, the variation in entropy production rates in Fig. \ref{Fig:S2} is more straightforward. For Fig. \ref{Fig:S2}(a), the profile of the variation in $\dot{S} _{\mathrm{NOEF}}^{\mathrm{A}} $ is almost identical to that in $\dot{S} _{\mathrm{NOEF}}^{\mathrm{A}} $ in the whole field, and $\dot{S} _{\mathrm{NOEF}}^{\mathrm{B}} $ is nonzero only when the shock wave passes through the mixing zone. For Fig. \ref{Fig:S2} (b), both $\dot{S} _{\mathrm{NOMF}}^{\mathrm{A}} $ and $\dot{S} _{\mathrm{NOMF}}^{\mathrm{B}} $ are generated primarily when the shock front impacts the substance interface, and there is little difference between the two substances. In Fig. \ref{Fig:S2} (c), $\dot{S} _{\mathrm{sum}}^{\mathrm{A}} > \dot{S} _{\mathrm{sum}}^{\mathrm{B}}$. The following conclusions are derived from a comparison of the entropy production rate profiles for the whole field and the mixing zone: (1) For the RMI and reshock processes studied in this paper, the mixing zone is the primary contributor to ${\dot{S} _{\mathrm{NOEF}} }$ throughout the entire flow process, whereas the flow field region excluding the mixing zone is the primary contributor to ${\dot{S} _{\mathrm{NOMF}} }$. (2) Whether RMI or reshock, the interaction between the shock wave and the substance mixing zone results in an increase in the entropy generation rates. However, the increase in entropy generation rates in reshock is smaller than that in the first shock because the shock intensity is weaker in reshock. (3) Substances A and B behave differently in terms of entropy production rates during the same process, and in this paper, the entropy production rate of substance A is higher than that of substance B.

\section{Conclusions and discussions}\label{sec:conclusions}

In this paper, we use a two-fluid DBM to simulate the Richtmyer-Meshkov instability and reshock processes and compare the results with experimental data. Our simulation results show good agreement with the experimental findings. In addition, a nonequilibrium kinetic analysis of the mixed substance interface in the RMI due to the initial shock and reshock processes is performed. Our findings show that the behavior of TNE quantities in the RMI case is markedly different from that in normal shock on an unperturbed plane interface between two uniform substances, which can be well described by the Rankine-Hugoniot relations. In the RMI case, the TNE quantities exhibit complex but fascinating kinetic effects both during and after the shock front.
The kinetic effects are described by two sets of TNE measures.
All four TNE measures in the first set abruptly increase in the shock process. The nonequilibrium quantity $\left |\bm{\Delta} _{3,1}^{ {\rm{*}}}\right |$, which corresponds to the heat flux(or NOEF), is caused by two factors: the temperature difference between the two substances and the shock wave effects. Both factors contribute to the heat flux to a similar extent, but for $\left |\bm{\Delta} _{2}^{ {\rm{*}}}\right |$, which corresponds to the viscous stress(or NOMF), the shock wave induces a much stronger nonequilibrium effect that is significantly larger than the $\left |\bm{\Delta} _{2}^{ {\rm{*}}}\right |$ generated by substance mixing. The behaviors of $\left |\bm{\Delta} _{3}^{ {\rm{*}}}\right |$ and $\left |\bm{\Delta} _{4,2}^{ {\rm{*}}}\right |$ are similar to those of $\left |\bm{\Delta} _{3,1}^{ {\rm{*}}}\right |$ and $\left |\bm{\Delta} _{2}^{ {\rm{*}}}\right |$, respectively. After the shock wave impacts the interface, it shows an ``instantaneous action'' for $D^{\rm{*T}}_{2}$ and $D^{\rm{*T}}_{4,2}$ but a ``continuous action'' for $D^{\rm{*T}}_{3,1}$ and $D^{\rm{*T}}_{3}$ in the mixing zone. When reshock occurs, the behaviors of $D^{\rm{*T}}_{3,1}$ and $D^{\rm{*T}}_{3}$ differ from those of $D^{\rm{*T}}_{2}$ and $D^{\rm{*T}}_{4,2}$ as well. In addition, the normalized nonequilibrium intensity quantity $D^{\rm{*T}}_{(4,2),\rm{nor}}$ in the mixing zone can be utilized as the sign of the interaction between the shock waves and substance interface.   Utilizing the two-fluid DBM, we evaluate the performance of the two substances. Due to their distinct properties, the two substances exhibit varying nonequilibrium behaviors when subjected to the same process. The behaviors of the flow system at the macroscopic level are determined by the collective responses of both substances. In addition, the distributions of TNE measures in the reshock process show informative patterns, which play a similar role to those obtained by the schlieren method used in shock wave experiments. The TNE quantities can show more flow field details.

The entropy production rates are the second set of TNE measures in the current work. Regarding the entropy production rate of each substance, the following results can be obtained: For the RMI and reshock processes studied in this paper, (i) the mixing zone is the primary contributor to ${\dot{S} _{\mathrm{NOEF}} }$, while the flow field region excluding the mixing zone is the primary contributor to ${\dot{S} _{\mathrm{NOMF}} }$. (ii) Since the shock wave is weaker in the reshock process, the increase in the entropy production rate during reshock is less than that observed during the initial shock. (iii) Each substance behaves differently in terms of entropy production rates, with the lighter fluid having a higher entropy production rate than the heavier fluid.

\section*{Acknowledgments}
The authors thank Feng Chen, Junfeng Wu, Chao Zhang, Zheng Yan, Jiahui Song, Dejia Zhang, Jie Chen, Hanwei Li, and Yingqi Jia for helpful discussions on the DBM. This work was supported by the National Natural Science Foundation of China (under Grant Nos. 12172061, 11875001, 11575033, and 11975053), the opening project of the State Key Laboratory of Explosion Science and Technology (Beijing Institute of Technology) (under Grant No. KFJJ23-02M), Foundation of National Key Laboratory of Shock Wave and Detonation Physics, and the Foundation of National Key Laboratory of Computational Physics.


\vspace*{2mm}
\begin{thebibliography}{86}
\vspace*{-1mm}
\begin{small}\baselineskip=10pt\itemsep-2pt
\bibitem{richtmyer1960taylor}R. D. Richtmyer, Taylor instability in shock acceleration of compressible fluids. Commun. Pur. Appl. Math. \textbf{13} (1960) 297.
\bibitem{meshkov1969instability}E. Meshkov, Instability of the interface of two gases accelerated by a shock wave. Fluid Dynam. \textbf{4} (1969) 101.
\bibitem{remington2006experimental}B. A. Remington, R. P. Drake, and D. D. Ryutov, Experimental astrophysics with high power lasers and z pinches. Rev. Mod. Phys. \textbf{78} (2006) 755.
\bibitem{nagel2022experiments}S. Nagel, K. Raman, C. Huntington, S. MacLaren, P. Wang, J. Bender, S. Prisbrey, and Y. Zhou, Experiments on the single-mode Richtmyer-Meshkov instability with reshock at high energy densities. Phys. Plasmas \textbf{29} (2022) 032308.
\bibitem{roycroft2022double}R. Roycroft, J. P. Sauppe, and P. A. Bradley, Double cylinder target design for study of hydrodynamic instabilities in multi-shell ICF. Phys. Plasmas  \textbf{29} (2022) 032704.
\bibitem{macphee2018hydrodynamic}A. MacPhee, V. Smalyuk, O. Landen, C. Weber, H. Robey, E. Alfonso, K. Baker, L. Berzak Hopkins, J. Biener, T. Bunn, et al., Hydrodynamic instabilities seeded by the X-ray shadow of ICF capsule fill-tubes. Phys. Plasmas  \textbf{25} (2018) 082702.
\bibitem{sauppe2019using}J. P. Sauppe, S. Palaniyappan, E. N. Loomis, J. L. Kline, K. A. Flippo, and B. Srinivasan, Using cylindrical implosions to investigate hydrodynamic instabilities in convergent geometry. Matter Radiat. Extrem. \textbf{4} (2019) 065403.
\bibitem{li2021growth}H. Li, B. Tian, Z. He, and Y. Zhang, Growth mechanism of interfacial fluid-mixing width induced by successive nonlinear wave interactions. Phys. Rev. E  \textbf{103} (2021) 053109.
\bibitem{xie2022chemical}Y. Xie, J. Shao, R. Liu, and P. Chen, Chemical reaction of Ni/Al interface associated with perturbation growth under shock compression. Phys. Fluids  \textbf{34} (2022) 044111.
\bibitem{li2021improved}Z. Li, W. Hu, J. Wu, and A. Peng, Improved gas-kinetic unified algorithm for high rarefied to continuum flows by computable modeling of the Boltzmann equation. Phys. Fluids \textbf{33} (2021) 126114.
\bibitem{liu2017molecular}H. Liu, Y. Zhang, W. Kang, P. Zhang, H. Duan, and X. He, Molecular dynamics simulation of strong shock waves propagating in dense deuterium, taking into consideration effects of excited electrons. Phys. Rev. E \textbf{95} (2017) 023201.
\bibitem{qiu2020study}R. Qiu, Y. Bao, T. Zhou, H. Che, R. Chen, and Y. You, Study of regular reflection shock waves using a mesoscopic kinetic approach: Curvature pattern and effects of viscosity. Phys. Fluids \textbf{32} (2020) 106106.
\bibitem{wu2022gaussian}Z. Chai, B. Shi, J. Lu, and Z. Guo, Non-Darcy flow in disordered porous media: A lattice Boltzmann study. Comput. Fluids \textbf{39} (2010) 2069-2077.
\bibitem{yang2022spatio}Z. Yang, C. Zhong, C. Zhuo, and S. Liu, Spatio-temporal error coupling and competition in meso-flux construction of discrete unified gas-kinetic scheme. Comput. Fluids \textbf{244} (2022) 105537.
\bibitem{jiang2019computation}Z. Jiang, W. Zhao, W. Chen, and R. Agarwal, Computation of shock wave structure using a simpler set of generalized hydrodynamic equations based on nonlinear coupled constitutive relations. Shock Waves \textbf{29} (2019) 1227.
\bibitem{CWF2016}Z. Chai, C. Huang, B. Shi, and Z. Guo, A comparative study on the lattice Boltzmann models for predicting effective diffusivity of porous media. Int. J. Heat Mass Tran. \textbf{98} (2016) 687-696.
\bibitem{li2015rarefied}Z. Li, A. Peng, H. Zhang, and J. Yang, Rarefied gas flow simulations using high-order gas-kinetic unified algorithms for Boltzmann model equations. Prog. Aerosp. Sci. \textbf{74} (2015) 81.
\bibitem{LC2019}C. Liu and K. Xu, Direct modeling methodology and its applications in multiscale transport process(in Chinese). Acta Aerodynamica Sinica \textbf{38} (2020) 197.
\bibitem{shan2006kinetic}X. Shan, X. Yuan, and H. Chen, Kinetic theory representation of hydrodynamics: a way beyond the Navier-Stokes equation. J. Fluid Mech. \textbf{550} (2006) 413.
\bibitem{meng2011accuracy}J. Meng and Y. Zhang, Accuracy analysis of high-order lattice Boltzmann models for rarefied gas flows. J. Comput. Phys. \textbf{230} (2011) 835.
\bibitem{chen2022simulation}T. Chen, X. Wen, L. Wang, Z. Guo, J. Wang, S. Chen, and D. B. Zhakebayev, Simulation of three-dimensional forced compressible isotropic turbulence by a redesigned discrete unified gas kinetic scheme. Phys. Fluids \textbf{34} (2022) 025106.
\bibitem{su2022temperature}W. Su, Q. Li, Y. Zhang, and L. Wu, Temperature jump and knudsen layer in rarefied molecular gas. Phys. Fluids \textbf{34} (2022) 032010.
\bibitem{guo2021progress}Z. Guo and K. Xu, Progress of discrete unified gas-kinetic scheme for multiscale flows. Advances in Aerodynamics \textbf{3} (2021) 1.
\bibitem{rinderknecht2018kinetic}H. G. Rinderknecht, P. Amendt, S. Wilks, and G. Collins, Kinetic physics in ICF: present understanding and future directions. Plasma Phys. Contr. F. \textbf{60} (2018) 064001.
\bibitem{shan2018experimental}L. Shan, H. Cai, W. Zhang, Q. Tang, F. Zhang, Z. Song, B. Bi, F. Ge, J. Chen, D. Liu, et al., Experimental evidence of kinetic effects in indirect-drive inertial confinement fusion hohlraums. Phys. Rev. Lett. \textbf{120} (2018) 195001.
\bibitem{cai2021hybrid}H. Cai, X. Yan, P. Yao, and S. Zhu, Hybrid fluid-particle modeling of shock-driven hydrodynamic instabilities in a plasma. Matter Radiat. Extrem. \textbf{6} (2021) 035901.
\bibitem{zhang2020species}S. Zhang and S. Hu, Species separation and hydrogen streaming upon shock release from polystyrene under inertial confinement fusion conditions. Phys. Rev. Lett. \textbf{125} (2020) 105001.
\bibitem{zhou2019turbulent}Y. Zhou, T. T. Clark, D. S. Clark, S. Gail Glendinning, M. Aaron Skinner, C. M. Huntington, O. A. Hurricane, A. M. Dimits, and B. A. Remington, Turbulent mixing and transition criteria of flows induced by hydrodynamic instabilities. Phys. Plasmas \textbf{26} (2019) 080901.
\bibitem{2017Rayleigh}Y. Zhou, Rayleigh-Taylor and Richtmyer-Meshkov instability induced flow, turbulence, and mixing. I. Phys. Rep. \textbf{720-722} (2017) 1.
\bibitem{zhou2017rayleigh}Y. Zhou, Rayleigh-Taylor and Richtmyer-Meshkov instability induced flow, turbulence, and mixing. II. Phys. Rep. \textbf{723} (2017) 1.
\bibitem{zou2020research}L. Zou, Q. Wu, and X. Li, Research progress of general Richtmyer-Meshkov instability. Sci. Sin-phys. Mech. As. \textbf{50} (2020) 104702.
\bibitem{zhang1996analytical}Q. Zhang and S. Sohn, An analytical nonlinear theory of Richtmyer-Meshkov instability. Phys. Lett. A \textbf{212} (1996) 149.
\bibitem{valerio1999modeling}E. Valerio, G. Jourdan, L. Houas, D. Zeitoun, and D. Besnard, Modeling of Richtmyer-Meshkov instability-induced turbulent mixing in shock-tube experiments. Phys. Fluids \textbf{11}  (1999) 214.
\bibitem{mikaelian1990rayleigh}K. O. Mikaelian, Rayleigh-Taylor and Richtmyer-Meshkov instabilities and mixing in stratified spherical shells. Phys. Rev. A \textbf{42}  (1990) 3400.
\bibitem{zhang2018quantitative}Q. Zhang, S. Deng, and W. Guo, Quantitative theory for the growth rate and amplitude of the compressible Richtmyer-Meshkov instability at all density ratios. Phys. Rev. Lett. \textbf{121} (2018) 174502.
\bibitem{liang2022phase}Y. Liang, The phase effect on the Richtmyer-Meshkov instability of a fluid layer. Phys. Fluids \textbf{34} (2022) 034106.
\bibitem{zhou2016asymptotic}Y. Zhou, W. H. Cabot, and B. Thornber, Asymptotic behavior of the mixed mass in Rayleigh-Taylor and Richtmyer-Meshkov instability induced flows. Phys. Plasmas \textbf{23} (2016) 052712.
\bibitem{lombardini2011atwood}M. Lombardini, D. Hill, D. Pullin, and D. Meiron, Atwood ratio dependence of Richtmyer-Meshkov flows under reshock conditions using large-eddy simulations. J. Fluid Mech. \textbf{670}  (2011) 439.
\bibitem{tritschler2014evolution}V. Tritschler, M. Zubel, S. Hickel, and N. Adams, Evolution of length scales and statistics of Richtmyer-Meshkov instability from direct numerical simulations. Phys. Rev. E \textbf{90} (2014) 063001.
\bibitem{ukai2011growth}S. Ukai, K. Balakrishnan, and S. Menon, Growth rate predictions of single-and multi-mode Richtmyer-Meshkov instability with reshock. Shock Waves \textbf{21} (2011) 533.
\bibitem{olson2014comparison}B. J. Olson and J. A. Greenough, Comparison of two-and three-dimensional simulations of miscible Richtmyer-Meshkov instability with multimode initial conditions. Phys. Fluids \textbf{26} (2014) 101702.
\bibitem{hahn2011richtmyer}M. Hahn, D. Drikakis, D. Youngs, and R. Williams, Richtmyer-Meshkov turbulent mixing arising from an inclined material interface with realistic surface perturbations and reshocked flow. Phys. Fluids \textbf{23} (2011) 046101.
\bibitem{schilling2010high}O. Schilling and M. Latini, High-order WENO simulations of three-dimensional reshocked Richtmyer-Meshkov instability to late times: dynamics, dependence on initial conditions, and comparisons to experimental data. Acta Math. Sci. \textbf{30} (2010) 595.
\bibitem{yan2022effect}Z. Yan, Y. Fu, L. Wang, C. Yu, and X. Li, Effect of chemical reaction on mixing transition and turbulent statistics of cylindrical Richtmyer-Meshkov instability. J. Fluid Mech. \textbf{941} (2022).
\bibitem{li2019role}H. Li, Z. He, Y. Zhang, and B. Tian, On the role of rarefaction/compression waves in Richtmyer-Meshkov instability with reshock. Phys. Fluids \textbf{31} (2019) 054102.
\bibitem{thornber2012energy}B. Thornber and Y. Zhou, Energy transfer in the Richtmyer-Meshkov instability. Phys. Rev. E \textbf{86} (2012) 056302.
\bibitem{li2018richtmyer}Y. Li, R. Samtaney, and V. Wheatley, The Richtmyer-Meshkov instability of a double-layer interface in convergent geometry with magnetohydrodynamics. Matter Radiat. Extrem. \textbf{3} (2018) 207.
\bibitem{zou2019richtmyer}L. Zou, M. Al-Marouf, W. Cheng, R. Samtaney, J. Ding, and X. Luo, Richtmyer-Meshkov instability of an unperturbed interface subjected to a diffracted convergent shock. J. Fluid Mech. \textbf{879} (2019) 448.
\bibitem{guan2017manipulation}B. Guan, Z. Zhai, T. Si, X. Lu, and X. Luo, Manipulation of three-dimensional Richtmyer-Meshkov instability by initial interfacial principal curvatures. Phys. Fluids \textbf{29} (2017) 032106.
\bibitem{sterbentz2022design}D. M. Sterbentz, C. F. Jekel, D. A. White, S. Aubry, H. E. Lorenzana, and J. L. Belof, Design optimization for Richtmyer-Meshkov instability suppression at shock-compressed material interfaces. Phys. Fluids \textbf{34} (2022) 082109.
\bibitem{jones1997membraneless}M. Jones and J. Jacobs, A membraneless experiment for the study of Richtmyer-Meshkov instability of a shock-accelerated gas interface. Phys. Fluids \textbf{9} (1997) 3078.
\bibitem{balasubramanian2013experimental}S. Balasubramanian, G. Orlicz, and K. Prestridge, Experimental study of initial condition dependence on turbulent mixing in shock-accelerated Richtmyer-Meshkov fluid layers. J. Turbul. \textbf{14} (2013) 170.
\bibitem{jacobs2013experiments}J. Jacobs, V. Krivets, V. Tsiklashvili, and O. Likhachev, Experiments on the Richtmyer-Meshkov instability with an imposed, random initial perturbation. Shock Waves \textbf{23} (2013) 407.
\bibitem{zhai2016richtmyer}Z. Zhai, P. Dong, T. Si, and X. Luo, The Richtmyer-Meshkov instability of a ``v" shaped air/helium interface subjected to a weak shock. Phys. Fluids \textbf{28} (2016) 082104.
\bibitem{weber2012turbulent}C. Weber, N. Haehn, J. Oakley, D. Rothamer, and R. Bonazza, Turbulent mixing measurements in the Richtmyer-Meshkov instability. Phys. Fluids \textbf{24} (2012) 074105.
\bibitem{vandenboomgaerde2014experimental}M. Vandenboomgaerde, D. Souffland, C. Mariani, L. Biamino, G. Jourdan, and L. Houas, An experimental and numerical investigation of the dependency on the initial conditions of the Richtmyer-Meshkov instability. Phys. Fluids \textbf{26} (2014) 024109.
\bibitem{balasubramanian2012experimental}S. Balasubramanian, G. Orlicz, K. Prestridge, and B. Balakumar, Experimental study of initial condition dependence on Richtmyer-Meshkov instability in the presence of reshock. Phys. Fluids \textbf{24} (2012) 034103.
\bibitem{liu2018elaborate}L. Liu, Y. Liang, J. Ding, N. Liu, and X. Luo, An elaborate experiment on the single-mode Richtmyer-Meshkov instability. J. Fluid Mech. \textbf{853} (2018).
\bibitem{luo2019nonlinear}X. Luo, M. Li, J. Ding, Z. Zhai, and T. Si, Nonlinear behaviour of convergent Richtmyer-Meshkov instability. J. Fluid Mech. \textbf{877} (2019) 130.
\bibitem{collins2002plif}B. Collins and J. Jacobs, Plif flow visualization and measurements of the Richtmyer-Meshkov instability of an air/SF$_6$ interface. J. Fluid Mech. \textbf{464} (2002) 113.
\bibitem{zou2017richtmyer}L. Zou, J. Liu, S. Liao, X. Zheng, Z. Zhai, and X. Luo, Richtmyer-Meshkov instability of a flat interface subjected to a rippled shock wave. Phys. Rev. E \textbf{95} (2017) 013107.
\bibitem{li2022instability}J. Li, J. Ding, X. Luo, and L. Zou, Instability of a heavy gas layer induced by a cylindrical convergent shock. Phys. Fluids \textbf{34} (2022) 042123.
\bibitem{ding2017shock}J. Ding, T. Si, J. Yang, X. Lu, Z. Zhai, and X. Luo, Measurement of a Richtmyer-Meshkov instability at an air/SF$_6$ interface in a semiannular shock tube. Phys. Rev. Lett \textbf{119} (2017) 014501.
\bibitem{lei2017experimental}F. Lei, J. Ding, T. Si, Z. Zhai, and X. Luo, Experimental study on a sinusoidal air/SF$_6$ interface accelerated by a cylindrically converging shock. J. Fluid Mech. \textbf{826} (2017) 819.
\bibitem{luo2018long}X. Luo, F. Zhang, J. Ding, T. Si, J. Yang, Z. Zhai, and C. Wen, Long-term effect of Rayleigh-Taylor stabilization on converging Richtmyer-Meshkov instability. J. Fluid Mech. \textbf{849} (2018) 231.
\bibitem{liang2019richtmyer}Y. Liang, Z. Zhai, J. Ding, and X. Luo, Richtmyer-Meshkov instability on a quasi-single-mode interface. J. Fluid Mech. \textbf{872} (2019) 729.
\bibitem{xu2012lattice}A. Xu, G. Zhang, Y. Gan, F. Chen, and X. Yu, Lattice Boltzmann modeling and simulation of compressible flows. Front. Phys. \textbf{7} (2012) 582.
\bibitem{Xu-Zhang-book-2022}A. Xu and Y. Zhang, Complex Media Kinetics (Science Press, Beijing, 2022), 1st ed.
\bibitem{gan2022discrete}Y. Gan, A. Xu, H. Lai, W. Li, G. Sun, and S. Succi, Discrete Boltzmann multi-scale modeling of non-equilibrium multiphase flows. J. Fluid Mech. \textbf{951} (2022) A8.
\bibitem{zhang2022discrete}D. Zhang, A. Xu, Y. Zhang, Y. Gan, and Y. Li, Discrete Boltzmann modeling of high-speed compressible flows with various depths of non-equilibrium. Phys. Fluids \textbf{34} (2022) 086104.
\bibitem{zhang2022non}Y. Zhang, A. Xu, F. Chen, C. Lin, and Z. Wei, Non-equilibrium characteristics of mass and heat transfers in the slip flow. AIP Adv. \textbf{12} (2022) 035347.
\bibitem{succi2001lattice}S. Succi, The lattice Boltzmann equation: for fluid dynamics and beyond (Oxford university press, 2001).
\bibitem{zhang2020kinetic}Y. Zhang, A. Xu, J. Qiu, H. Wei, and Z. Wei, Kinetic modeling of multiphase flow based on simplified enskog equation. Front. Phys. \textbf{15} (2020) 1.
\bibitem{sun2022}G. Sun, Y. Gan, A. Xu, Y. Zhang, and Q. Shi, Thermodynamic non-equilibrium effects in bubble coalescence: A discrete Boltzmann study. Phys. Rev. E \textbf{106} (2022) 035101.
\bibitem{zhang2018discrete}Y. Zhang, A. Xu, G. Zhang, and Z. Chen, Discrete Boltzmann method with Maxwell-type boundary condition for slip flow. Commun. Theor. Phys. \textbf{69} (2018) 77.
\bibitem{Lin2016CNF}C. Lin, A. Xu, G. Zhang, and Y. Li, Double-distribution-function discrete Boltzmann model for combustion. Combust. Flame \textbf{164} (2016) 137.
\bibitem{Lin2018CNF}C. Lin and K. H. Luo, Mesoscopic simulation of nonequilibrium detonation with discrete Boltzmann method. Combust. Flame \textbf{198} (2018) 356.
\bibitem{shan2022discrete}Y. Shan, A. Xu, Y. Zhang, L. Wang, and F. Chen, Discrete Boltzmann modeling of detonation: Based on the Shakhov model. P. I. Mech. Eng. C-J. Mec. p. \textbf{237} (2023) 2517-2531.
\bibitem{Lai2016PRE}H. Lai, A. Xu, G. Zhang, Y. Gan, Y. Ying, and S. Succi, Nonequilibrium thermohydrodynamic effects on the Rayleigh-Taylor instability in compressible flows. Phys. Rev. E \textbf{94} (2016) 023106.
\bibitem{Gan2019PRE}D. Zhang, A. Xu, J. Song, Y. Gan, Y. Zhang, and Y. Li, Specific-heat ratio effects on the interaction between shock wave and heavy-cylindrical bubble: Based on discrete Boltzmann method. Comput. Fluids \textbf{265} (2023) 106021.
\bibitem{chen2022}J. Chen, A. Xu, D. Chen, Y. Zhang, and C. Zhihua, Discrete Boltzmann modeling of Rayleigh-Taylor instability: Effects of interfacial tension, viscosity, and heat conductivity. Phys. Rev. E \textbf{106} (2022) 015102.
\bibitem{latini2007high}M. Latini, O. Schilling, and W. S. Don, High-resolution simulations and modeling of reshocked single-mode Richtmyer-Meshkov instability: Comparison to experimental data and to amplitude growth model predictions. Phys. Fluids \textbf{19} (2007) 024104.
\bibitem{zel2002physics}Y. B. Zel'Dovich and Y. P. Raizer, Physics of shock waves and high-temperature hydrodynamic phenomena (Courier Corporation, 2002).
\bibitem{zhang2019entropy}Y. Zhang, A. Xu, G. Zhang, Y. Gan, Z. Chen, and S. Succi, Entropy production in thermal phase separation: a kinetic-theory approach. Soft matter \textbf{15} (2019) 2245-2259.
\bibitem{youngs1991three}D. L. Youngs, Three-dimensional numerical simulation of turbulent mixing by Rayleigh-Taylor instability. Phys. Fluids A: Fluid Dynamics \textbf{3} (1991) 1312.
\bibitem{youngs1994numerical}D. L. Youngs, Numerical simulation of mixing by Rayleigh-Taylor and Richtmyer-Meshkov instabilities. Laser Part. Beams \textbf{12} (1994) 725.

\end{small}
\end{thebibliography}
\end{document}